\documentclass[conference]{IEEEtran}
\IEEEoverridecommandlockouts
\usepackage{caption}
\usepackage{cite}
\usepackage{amsmath,amssymb,amsfonts}
\usepackage{algorithmic}
\usepackage{graphicx}
\usepackage{textcomp}
\usepackage{xcolor}
\usepackage{hyperref}
\usepackage[disable]{todonotes}
\usepackage{makecell}
\usepackage{xcolor}
\usepackage{tabularx}
\usepackage[most]{tcolorbox}
\usepackage{orcidlink}
\usepackage{hhline}
\usepackage{multirow}
\usepackage{colortbl}
\usepackage{soul}

\definecolor{role}{HTML}{85645F}
\definecolor{cot}{HTML}{B34FE2}
\definecolor{emotional}{HTML}{CD4E18}
\definecolor{variable}{HTML}{3E89D2}
\newcommand{\python}{Python}
\newcommand{\solver}{SMT}
\newcommand{\hybrid}{Hybrid}

\newcommand{\gsr}{GSR}
\newcommand{\precision}{P}
\newcommand{\recall}{R}
\newcommand{\fscore}{F1}

\newcommand{\llm}[1]{\textbf{#1}}

\newcommand{\ind}{\texttt{IND}}
\newcommand{\batch}{\texttt{BATCH}}

\newcommand{\set}[1]{\mathcal{#1}}

\newcommand{\removed}[1]{{\color{red}\st{#1}}}
\newcommand{\added}[1]{{\color{blue}#1}}
\newcommand{\reviewid}[1]{{\color{green}#1}}

\renewcommand{\removed}[1]{}
\renewcommand{\added}[1]{#1}
\renewcommand{\reviewid}[1]{}

\def\BibTeX{{\rm B\kern-.05em{\sc i\kern-.025em b}\kern-.08em
    T\kern-.1667em\lower.7ex\hbox{E}\kern-.125emX}}
\begin{document}

\title{
% LLM-based Requirements Verification Through Consistent Text Data Generation
% Consistent Textual Data Generation for Requirements-based Testing
% \thanks{\textsuperscript{*}Both authors contributed equally to the research.}
% LLM-based Requirements Verification Through Consistent Text Data and Checker Generation
LLM-based Satisfiability Checking of  \\
String Requirements by Consistent Data and \\
Checker Generation
\thanks{\textsuperscript{*}Both authors contributed equally to the research.}
\thanks{\textsuperscript{†}Also affiliates with McGill University.}
\thanks{
Partially supported by the FRQNT-B2X project (file number: 319955), IT30340 Mitacs Accelerate, and the Wallenberg AI, Autonomous Systems
and Software Program (WASP), Sweden
}
}

\author{
\IEEEauthorblockN{Boqi Chen\textsuperscript{*}\orcidlink{0000-0002-1451-3603}}
\IEEEauthorblockA{\textit{ECE} \\
\textit{McGill University}\\
Montreal, Canada}
\and
\IEEEauthorblockN{Aren A. Babikian\textsuperscript{*}\orcidlink{0000-0002-8108-0043}}
\IEEEauthorblockA{\textit{Dep't of Computer Science} \\
\textit{University of Toronto}\\
Toronto, Canada}
\and
\IEEEauthorblockN{Shuzhao Feng\orcidlink{0009-0001-0743-7219}}
\IEEEauthorblockA{\textit{ECE} \\
\textit{McGill University}\\
Montreal, Canada}
\and
\IEEEauthorblockN{D\'aniel Varr\'o\textsuperscript{†} \orcidlink{0000-0002-8790-252X}}
\IEEEauthorblockA{
\textit{IDA} \\
\textit{Link\"oping University}\\
Link\"oping, Sweden}
\and
\IEEEauthorblockN{Gunter Mussbacher\orcidlink{0009-0006-8070-9184}}
\IEEEauthorblockA{\textit{ECE} \\
\textit{McGill University}\\
Montreal, Canada}
}

\maketitle
% \todo{Ensure the correct use of "requirement" vs. "constraint"}
% \todo{Consistent vs. satisfiable vs. feasible}
% \todo{confirm title}
% \todo{text vs. string}

\begin{abstract}

% To facilitate collaboration among stakeholders with diverse expertise, requirements are often represented using natural language (NL).
Requirements over strings, commonly represented using natural language (NL), are particularly relevant for software systems due to their heavy reliance on string data manipulation.
While individual requirements can usually be analyzed manually, verifying properties (e.g., satisfiability) over sets of NL requirements is particularly challenging.
Formal approaches (e.g., SMT solvers) may efficiently verify such properties, but are known to have theoretical limitations.
Additionally,
% However,
the translation of NL requirements into formal constraints typically requires significant manual effort. 
% Furthermore, such approaches have known limitations for tasks governed by an undecidable theory.
Recently, large language models (LLMs) have emerged as an alternative approach for formal reasoning tasks, but their effectiveness in verifying requirements over strings is less studied.
%
% In this paper, we introduce a hybrid generative approach to verify the satisfiability of NL requirements over text.
In this paper, we introduce a hybrid approach that verifies the satisfiability of NL requirements over strings by using LLMs (1) to derive a satisfiability outcome (and a consistent string, if possible), and (2) to generate declarative (i.e., SMT) and imperative (i.e., Python) checkers, used to validate the correctness of (1). 
% if a text value satisfies each individual requirement.
% It then uses these checkers to provides either valid text data as a proof of satisfiability or detects unsatisfiable NL requirements.
% The generated text data can subsequently be used to verify whether a system meets these requirements.
%
In our experiments, we assess the performance of four LLMs.
% by integrating two types of checkers for text data.
Results show that LLMs effectively translate natural language into checkers, even achieving perfect testing accuracy for Python-based checkers. 
These checkers substantially help LLMs in generating a consistent string and accurately identifying unsatisfiable requirements, leading to more than doubled generation success rate and F1-score in certain cases compared to baselines without generated checkers.

\end{abstract}
% \todo{maybe replace "specification-based testing" by "data generation"
% }


\section{Introduction}
\label{sec:introduction}

% \todo[inline]{the intro now focusses on verifying SOME property on requirements. Not specifically verifying satisfiability. I like this.}

% \todo[inline]{One fundamental question I have. Do we want to keep "data generation" as a main focus of the paper? (e.g., in the title). Reading through the intro, data generation is really looking like only a side-effect of our approach, and the core component is the checker. I think checkers are significantly more important for our approach, and we should not highlight data generation at the same level of checkers in the title.}

\textbf{Motivation.}
Large-scale software systems commonly involve numerous parties, from domain experts to technical team members and end-users.
For the successful development of such systems, it is necessary to ensure a smooth interaction between these parties, which becomes particularly challenging when considering their varying levels of domain expertise and technical knowledge.
For this purpose, requirements are most commonly captured using natural language (NL).
% from a requirements engineering perspective, it has become customary to represent system requirements using natural language (NL).

A particularity of software systems is their reliance on strings, not only for the implementation of software artifacts (e.g., through textual programs) but also for data manipulation. 
For instance, many systems manipulate strings in the form of an email address and a password given by the user as input. 
Additionally, strings must comply with well-formedness requirements (e.g., an email address must contain exactly one ``@'' character), which are often numerous and interconnected.
% As such, while it is often feasible to verify properties (e.g. satisfiability) of individual requirements via manual analysis, this becomes particularly challenging when considering sets of requirements over text data.\
Verifying the satisfiability of an individual requirement over strings may be feasible via manual analysis.
However, with an increasing number of requirements, common for large-scale projects, verifying their satisfiability becomes a challenging task, particularly considering that requirements may be contradictory or ambiguous. Such challenges have been studied by existing work in the field of requirements engineering~\cite{feng2024normative,fantechi2023inconsistency,hosseiniAmbiguityGeneralityNatural2021,giannakopoulouGenerationFormalRequirements2020,berryAmbiguityRequirementsSpecification2004}.
% the verification of their satisfiability becomes particularly challenging, considering that requirements may be contradictory or ambiguous.
In fact, inconsistency and ambiguity detection in (formal) requirements, even beyond strings, is an active subfield of requirements engineering.
\textbf{Problem Statement.}
Existing approaches for satisfiability checking of requirements over strings are often formal: they rely on general-purpose SMT solvers, such as Z3 \cite{de2008z3}, or on string-specific constraint solvers, such as Ostrich \cite{chenDecisionProceduresPath2019}.
Such approaches are sound and efficiently handle complex constraints over strings by either (1) detecting problems, i.e., unsatisfiability, in input requirements or (2) generating conforming strings as proof of satisfiability, which is particularly useful in the context of requirements-based testing.
However, \textit{a major assumption} of such approaches is to take a formal definition of requirements as input (e.g., using the SMT-LIB2 syntax), which is often unavailable (and hard to create by formal methods non-experts).
% In the context of requirement-based testing of textual artifacts, test cases require textual data that conforms to, or intentionally violates, system requirements, which define constraints over strings.
% To derive such constraint-satisfying string data, existing approaches are often formal: they rely on general-purpose SMT solvers such as Z3 \cite{de2008z3} or on string-specific constraint solvers, such as Ostrich \cite{chenDecisionProceduresPath2019}.
% While such approaches are sound and well-suited for handling complex string constraints efficiently, \textit{a major drawback} is that they rely on a formal definition of requirements (e.g. using the SMT-LIB2 syntax), which is often unavailable (and hard to create by formal methods non-experts).

% In many software systems, requirements are instead represented in natural language.
% Given that requirements are dominantly written in NL, approaches that support NL as input are preferred in order to achieve more generalizable and impactful results. 
Given that string requirements are mainly written in NL, NL-based approaches are better suited for their interpretation.
In recent years, large language models (LLMs) have emerged as a promising alternative for NL interpretation in general, with certain logic reasoning capabilities \cite{huang2023towards}, particularly due to their strong few-shot generalization capabilities in data-scarce settings.
However, existing research \cite{babikian2025exploring} has shown that the effectiveness of using LLMs to verify the satisfiability of (NL) string requirements is below that of formal approaches.
Still, their inherent support for NL requirements, as well as their capacity of generating \textit{realistic} strings as proof of consistency, remain \textit{significant assets} vis-\`{a}-vis formal approaches.

% \todo[inline, color=yellow]{New content below, please check!}
A further drawback of LLM-based verification of requirements satisfiability is the lack of soundness guarantees.
% that generated strings indeed satisfy the input constraints.
To remedy this, existing LLM-based approaches \cite{li2024guiding,first2023baldur,kirchner2024prover} rely on a trial-and-error method guarded by formal checks: LLM-derived outcomes are formally checked for correctness at each iteration and may only be returned if the check succeeds.
On the one hand, checking may be performed using (declarative) SMT solvers, which have known theoretical limitations, e.g., supporting nested quantifiers, which do occur in problems over strings.
% For instance, if, for instance, the checking task is governed by an undecidable theory.
For instance, checking if a string is a palindrome is complex since it requires the solver to check that \textit{for each} character at position $i$ of data $d$, there \textit{exists} an identical character at position $d.length-i$. 
On the other hand, imperative methods such as Python programs may efficiently check that strings conform to complex requirements, with less theoretical limitations.
However, they cannot be used to provide claims that the requirements are unsatisfiable.

% Such validation may be performed efficiently using imperative methods, e.g. Python functions.
% However, impeerative methods cannot be used in cases where LLM output validation requires evaluating the (un)satisfiability of the input constraints set (e.g. if the LLM claims that the constraints are unsatisfiable).
% Such cases may be addressed by (declarative) SMT solvers, which have known theoretical limitations (compared to imperative verification approaches) if, for instance, the verification task is governed by an undecidable theory.

% Pytho used for verification because extremely expressiveandfor correctness using output is verified for correctness  often rely on a 
% fragments of FOL w/ quantifiers
% undecideable 

% \todo{Add references saying that cyber-security is the main motivation for string solving}
% \todo{Add references that string-based analysis, e.g. mutation of characters, is relevant for testing or repairing string artifacts}

%%%%%%%%%%%%%%%%%%%%%%%%%%%%%%%
% RANDOM IDEA about DSL testing

% \todo{generate XML, JSON, hashcode}

% \todo{introduce the connnection between NL constraint + constraint "write }

% We ar ebetter because we have an informal definition of a DSL. A rough idea. can we generate string from this

% Top lvl = informal specificationo of DSL
% mid lvl = NL rep of each condition
% low lvl = SMT constraint

% testing DSL parser (future work)

\textbf{Contributions:}
% In this paper, we propose a hybrid approach relying on LLMs, SMT solvers, and imperative programming for the verification of NL requirements through the generation of consistent text data and checkers.
We introduce a hybrid generative approach to verify the satisfiability of NL requirements over strings.
Our approach uses LLMs (1) to derive a satisfiability outcome,
% (i.e., SAT + valid text as proof, or UNSAT),
and (2) to generate checkers used to ensure the correctness of the satisfiability outcome derived in (1).
Our approach generates two types of checkers:  declarative checkers (i.e., SMT specifications to verify unsatisfiability of requirement sets) and imperative checkers (i.e., Python programs, which can more generally check data when a requirement is satisfiable).
% string data generation in the context of requirement-based testing.
% Our approach leverages the leveraging SMT solvers and imperative programs, f
% Our approach takes as input a collection of natural-language constraints over strings and yields a conforming (or intentionally non-conforming) string as output.
The specific contributions of the paper are as follows:
% \todo[inline, color=yellow]{Terminaology question: we use data instead of strings. Should we use outcome instead of data? to include the UNSAT outcome possibility?}
\begin{itemize}
    \item We propose a novel approach that takes as input a collection of NL requirements over strings and either (1) detects unsatisfiable requirements or (2) generates a conforming string (instance) as a proof of satisfiability. (\autoref{sec:overview})
    \item We integrate a divide-and-conquer approach for efficient handling of complex collections of string requirements.
    % \todo{If we consider the divide-and-conquer aspect as a contributions, then we will probably need to compare it to a all-in-one approach, and call the all-in-one approach a baseline. For this we can maybe refer to MO2RE}
    As a novel intermediate step, our approach derives checkers for verifying the satisfiability of individual NL requirements and leverages a formal solver within a feedback loop to improve accuracy. (\autoref{sec:loop1})
    % \todo[color=yellow]{TODO Do we want to talk about the agent interactions?}
    % \item Refined approach to derive test data from set of input requirements over (structured or unstructured) strings
    % \item DISCUSS: mention some genetic algorithm here? This would make the approach very wide-scoped, which is great. Or with this, we may target wider-scope conferences in the future (e.g., ICSE)
    % \item DISCUSS: Inconsistency detection and traceability in input requirements (can we, for example, find inconsistency in catalog of SQL hazards?)
    % \item DISCUSS: do we want to do actual testing of systems? (SQL faults, some random system from a course)
    \item
    % Our approach relies on smart usage of SMT solvers and imperative programs to ensure the soundness of generated test data.
    Our approach relies on checkers to examine the correctness of LLM-derived satisfiability outcomes.
    % checker-based validation of the LLM-derived verification outcome
    % to provide soundness guarantees relative to checker correctness.
    In the case where no correct outcome is derived, our approach degrades gracefully by returning the closest-to-sound outcome derived within the allocated budget. (\autoref{sec:loop2})
    \item We conduct experiments with four popular LLMs, (i.e., GPT-4o \cite{hurst2024gpt}, GPT-4o-mini \cite{hurst2024gpt}, Llama3.1-8b \cite{grattafiori2024llama}, and DeepSeek-V3 \cite{liu2024deepseek}).
    % and integrate three state-of-the-art formal approaches (i.e. Z3 \cite{de2008z3}, Z3str \cite{berzish2017z3str3}, cvc5 \cite{barbosa2022cvc5}).
    We investigate 340 NL requirement sets conforming to 12 categories of textual variables drawn from common programming exercises in technical interviews.
    % undergraduate-level software engineering course projects.
    % We assess constraints at varying levels of complexity. 
    (\autoref{sec:evaluation})
\end{itemize}

\textbf{Added Value:}
In contrast to approaches focused solely on data generation, our approach provides accurate outcomes for both satisfiable and unsatisfiable (string) requirement sets.
% In contrast to recent results \cite{babikian2025exploring}, our approach focuses on providing accurate outcomes for both satisfiable and unsatisfiable text requirement sets.
% This paper builds on recent results \cite{babikian2025exploring} assessing the capacity of LLMs for string-constraint solving, which is a subset of NL requirements verification over text.
As a key improvement, this paper generates verification outcomes by taking as input \textit{only} the NL requirements, i.e., without the need to manually formalize NL requirements.
Additionally, our comparative evaluation gives insights on the strengths and weaknesses of various LLMs and checker types in providing correct verification outcomes for NL requirements.
% Additionally, we provide insights on the usability of the generated  proof-of-consistency text data in the context of requirements-based testing.

% diff LLMs
% how to gen checkers (batch vs 1)
% gen out: what feedback info
% gen out: what kind of checkers

% \todo[color=yellow]{Section for Structure, if necessary}
% \textbf{Structure:}

% For instance, to avoid using SQL queries vulnerable to fault injection,  when considering SQL queries as (textual) strings, 
% An example of such a rule is:
% While such a rule is 
% \todo{How is this not a syntactic constraint? Maybe I should change the example from the previous paragraph on the syntax of the programming langauge? Not sure.}
% strings are often the silent enabler of software vulnerabilities,
%  and a bad string manipulation can have disastrous effects especially for web applications developed in languages like
%  PHP or JavaScript.
\section{Background}

% \begin{itemize}
%     % \item Motivating case study: Should probably be for some SQL hazard-related stuff. Structurally, it can be very similar to the BDD stuff from MO2RE
%     \item Constraint satisfaction problem over strings (same as MO2RE)
%     \item LLMs . In here maybe add some stuff about feedback, multi-agent discussion, etc.
% \end{itemize}

\subsection{Satisfiability checking of requirements over strings}
\label{sec:back-reqVer}

% This paper focuses on generating textual test input data from requirements given as input.
This paper focuses on verifying the satisfiability of requirements (given as input) over strings by leveraging a consistent string data and checker generation approach.
To illustrate this, let us consider, as an illustrating example, an extended version of the email address manipulation system introduced in \autoref{sec:introduction} which has the following three requirements:
% only allows for inputs satisfying the following three requirements:
\begin{enumerate}
    \item [R1] The system shall require an email address to contain exactly one ``@'' character.
    \item [R2] The system shall require an email address not to end with a ``.'' character.
    \item [R3] The system shall require an email to contain at least one ``.'' character after the ``@'' character.
    % \item [R3] If the email contains a "@" character, then the email shall include a ".“ character after the "@” character but before the end.
\end{enumerate}

We illustrate the versatility of requirements satisfiability verification by considering the above system in a requirements-based testing context.
Given requirements $R1$, $R2$, and $R3$, a test engineer first selects a set of \textit{test cases}, represented as conditions over the set of requirements.
They then associate an \textit{expected system behavior} to each test 
case (e.g., the system stores the email address if and only if the given input satisfies all system requirements).
% case: correct behavior (e.g., the system stores the email address) is expected from the system if and only if the given input satisfies all system requirements.
An example containing five such test cases is shown in \autoref{tab:background table}.

% \begin{table}[htp]
%     \caption{Test cases and corresponding (textual) test data derived for }
%     \label{tab:background table}
%     \centering
%     \begin{tabular}{c|c|c}
%         Test cases & Consistency & Test data\\
%         \hline
%         $R1 \wedge R2 \wedge R3$ & Store email in database & "ab@cd.com"\\
%         $\neg R1 \wedge R2 \wedge R3$ & Show error message & "ab@@cd.com"\\
%         $R1 \wedge \neg R2 \wedge R3$ & Show error message & "ab@cd."\\
%         $R1 \wedge R2 \wedge \neg R3$ & Show error message & "ab@cdcom"\\
%         $R1 \wedge \neg R2 \wedge \neg R3$ & \textit{UNSAT} & \textit{UNSAT}\\
%     \end{tabular}
% \end{table}

\begin{table}[htp]
    \caption{Test cases, their satisfiability, and corresponding test data}
    \label{tab:background table}
    \centering
    \begin{tabular}{c|c|c|c}
        Test case & \makecell{Expected\\Behavior} & Satisfiability & Test data\\
        \hline
        $R1 \wedge R2 \wedge R3$ & Store email & Yes & ``ab@cd.com''\\
        $\neg R1 \wedge R2 \wedge R3$ & Show error & Yes & ``ab@@cd.com''\\
        $R1 \wedge \neg R2 \wedge R3$ & Show error & Yes & ``ab@cd.''\\
        $R1 \wedge R2 \wedge \neg R3$ & Show error & Yes & ``ab@cdcom''\\
        $R1 \wedge \neg R2 \wedge \neg R3$ & Show error & No & \textit{-}\\
    \end{tabular}
\end{table}

Upon test case selection, as a first step, the engineer must check whether the test case is satisfiable,
% (i.e., verify its consistency),
which is a challenging task to perform manually, even for domain experts.
For instance, it is difficult to determine manually that a $R1 \wedge \neg R2 \wedge \neg R3$ test case is unsatisfiable, considering that if an email containing exactly one ``@'' character ($R1$) does not contain any ``.'' characters after the ``@'' ($\neg R3$), then it cannot end with a ``.'' ($\neg R2$).
While SMT solvers may efficiently detect such an unsatisfiable case if the requirements are formalized, a key advantage of our approach is to derive \textit{SAT}/\textit{UNSAT} outcomes directly from NL requirements.

Once the satisfiability is determined, for each satisfiable test, the engineer must derive \textit{textual test data} (i.e., a string) that would trigger the expected behavior in the underlying system.
Deriving such data manually and ensuring its compliance with the test case specification is a daunting task, particularly with an increasing number of requirements.
Our approach generates conformant test data automatically for each satisfiable test case as a side-effect of the satisfiability verification, where the generated test data is the proof of satisfiability.
% In our paper, we provide an automated approach for requirements verification

% Then, for all feasible test cases, he must define
% % (either manually or using an automated approach such as the one proposed in this paper)
% (textual) test data that would trigger the appropriate function of the underlying system.
% With an increased number of requirements, not only does the number of relevant test cases increase but so does the complexity of the requirement verification and text data generation challenge.
% Luckily, our approachdoes this natively

% The problem becomes particularly challenging when considering the existence of string-specific corner cases as part of the test suite.
% For instance, it is difficult to manually understand that a $R1 \wedge \neg R2 \wedge \neg R3$ test case is unsatisfiable, since a string cannot end with a "." ($\neg R2$) if it does not contain any "." characters after "@" ($\neg R3$).
% To address this challenge, we leverage automated approaches for generating consistent strings (by LLMs), verifying  the correctness of the synthesized strings (by SMT solvers and imperative programs), and detecting unsatisfiable corner cases (by SMT solvers).

\subsection{Constraint satisfaction problems over strings}
Constraints over strings are common in the context of commercial software applications, e.g., through password and email restrictions.
They are also addressed significantly in scientific research, through automated, string-specific formal reasoning techniques \cite{chenDecisionProceduresPath2019,chenSolvingStringConstraints2023,lotzSolvingStringConstraints2023,wuDecisionProcedureString2024,kringsConstraintLogicProgramming2020} designed to solve collections of complex constraints.
These approaches are evaluated over large benchmarks \cite{SMTLIBSatisfiabilityModulo} 
% \footnote{\url{https://smt-lib.org/benchmarks.shtml}}
and in the context of solver competitions \cite{SMTCOMP2024},
% \footnote{\url{https://smt-comp.github.io/2024/}}
where string constraints are represented formally (e.g., in the SMT-LIB2 \cite{barrettSMTLIBStandardVersion2017} format).
In this context, the solvers are evaluated according to their achieved runtime and success rate.

While the constraint satisfaction problems contained in these benchmarks are valuable from a logical perspective, their relevance for real-world application domains such as requirements engineering has been less studied.
% Non-functional parameters, such as qualitative analysis (e.g. realism, relevance) of the generated string, are not considered, which limits the extensibility of formal solvers to domains such as requirement engineering.
Additionally, solvers take as input a formal representation of the problem, which is often unavailable upfront for most software systems.

\subsection{Large language models (LLMs)}
% LLMs are advanced neural networks based on the transformer architecture \cite{vaswani2017attention}, initially designed for language modeling tasks.
% \todoab{Percy, can you reword the sentence above?}
LLMs are powerful transformer neural networks \cite{vaswani2017attention} trained on vast amounts of data for handling various tasks involving natural language. 
% Through instruction fine-tuning, LLMs can be trained to follow input instructions, enabling them to adapt to diverse tasks via \textit{in-context learning} using a few examples or even zero-shot predictions with only task descriptions. 
Recent research highlights LLMs' emerging reasoning capabilities, achieving notable performance on many logic reasoning benchmarks \cite{huang2023towards}. This positions them as promising alternatives to traditional constraint solvers for addressing constraint-solving problems involving text values. 

% Since LLM takes natural language as input, various studies have been conducted on how to prompt the LLM for better performance. This trend leads to an actively studied field called prompt engineering. Among them, zero-shot chain-of-though reasoning \cite{kojima2022large} and emotional stimuli \cite{li2023large} have been shown effective across various LLMs. 

Recent research has also focused improving LLM performance by optimizing the input NL prompt, known as \textit{prompt engineering} \cite{sahoo2024systematic}.
% Since LLMs accept natural language as input, a significant amount of research has been conducted to optimize prompts for improving performance, an active area know as \textit{prompt engineering}. 
Techniques such as leveraging emotional stimuli \cite{li2023large} and zero-shot chain-of-thought reasoning \cite{kojima2022large} have been shown to be effective across various LLMs. 

While LLMs typically produce natural language output, recent advancements such as Open AI's \textit{structured outputs} \cite{openai2025introducing} and LangChain's \textit{output parsers} \cite{harrison2022langchain} have enabled them to produce structured output. In this paper, we rely on LangChain to instruct the LLM to present final answers in various structured formats depending on the components. 

% \todo{There is a section here about constraint complexity which we probably will not need}
% \subsection{Complexity of constraints}
% CHomsky hierarchy
% how these fit within different logic theories (P, NP, (un)decideable)

% SMT is designed to check sat/unsat of set of constraints.
% As a side-effect, in sat case, it can give an example.
% and can check correctness of variable value given as input.
% However, has limitation for certain undecideable theories.
% give example inn textual domain

% imperastive approaches cannot give sat/unsat. and cannot generate data.
% However can easily check correctness of input. and dont have same limitation.
% given same example, show that its easy to write a python function to check that

% in this paper, we leverage strength of each approach. for UNSAT verif, we can only do SMT
% for example verif, we can do both. we evluuate both

\section{Approach Overview}
\label{sec:overview}

% \todo[inline]{Note for Gunter and Daniel: We removed the test case generation entirely to better fit the requirements verification positioning}

\begin{figure}[htp]
    \centering
    \includegraphics[width=0.8\linewidth]{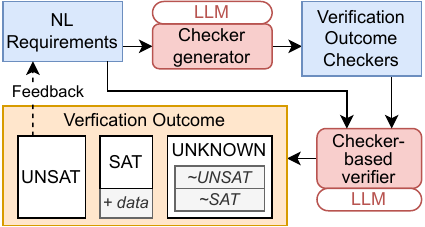}
    \caption{Approach overview}
    \label{fig:overview}
\end{figure}

An overview of our proposed satisfiability verification approach for NL requirement is shown in \autoref{fig:overview}, where the blue and orange boxes represent input, intermediate, and output artifacts, while the red boxes represent processes. Our approach takes as input a conjunction of \textit{NL requirements} 
% $R_1, \dots, R_n$
such as the ones introduced in \autoref{sec:back-reqVer}.
% Note that input requirements may be either positive or negative.
Our approach outputs a \textit{Verification Outcome} $v_i \in \{\textit{UNSAT}, \textit{SAT}, \textit{UNKNOWN}\}$ assessing the satisfiability of the input requirements conjunction.
% intermediate NL test case derived from the NL requirements.

% As a first step, our approach derives \textit{NL test cases} from the input requirements. Test cases are represented as Boolean formulae over the requirements, as in \autoref{tab:background table}. This allows us to formally verify the consistency of such test cases given the satisfiability of each individual requirement.
% Test case derivation from NL requirements (e.g. deriving a test suite that satisfies certain code coverage criteria given as input) is outside the scope of this paper, thus we generate test cases manually.
% In this paper, NL test case derivation is done manually.
% In practice, , in accordance to testing requirements and best practices observed by the test engineer.
% For instance, (requirements-based) testing safety-critical systems often requires the creation of an (NL) test suite that satisfies coverage criteria such as MC/DC coverage.

As a first step, our approach generates a \textit{Verification Outcome Checker} (checker) for each input NL requirement.
A checker is a process that takes as input (1) the NL requirements and (2) an intermediate \textit{UNSAT} or \textit{SAT} verification outcome for the requirements derived by an intermediate, not-necessarily-sound approach (e.g., LLMs), as detailed in \autoref{sec:loop2}.
It then checks (validates) the correctness of the intermediate verification outcome wrt. the requirements.
% The checker types may vary depending on the domain.
% any formal approach, e.g. for domain-specific applications, in accordance to the addressed verification task.
The \textit{checker generator} component is described in \autoref{sec:loop1}.

NL requirements and the generated checkers are then given as input to the \textit{checker-based verifier}, detailed in \autoref{sec:loop2}, which outputs a verification outcome for the conjunction of input NL requirements.
When the requirements are determined to be satisfiable, the verifier returns \textit{SAT}, as well as \textit{string data} as a proof of satisfiability.
When unsatisfiable NL requirements are detected,
% this shows that the corresponding boolean formula over the NL requirements is incon
the verifier returns \textit{UNSAT} and may pinpoint the problematic subset of the NL requirements through a \textit{feedback} loop.
\todo{R3-1}\added{Similar to existing heuristic-based search approaches \cite{dinges2024solving}, a budget is imposed on the feedback loop to limit the number of retries.}
If the verifier runs out of budget, it returns an \textit{UNKNOWN} outcome.
As a key particularity, our approach degrades gracefully when no sound \textit{SAT} or \textit{UNSAT} outcome may be provided: in this case, our approach returns the closest-to-sound outcome derived within the allocated budget.

\section{Checker Generator}
\label{sec:loop1}

\begin{figure*}[htp]
    \centering
    \includegraphics[width=0.8\linewidth]{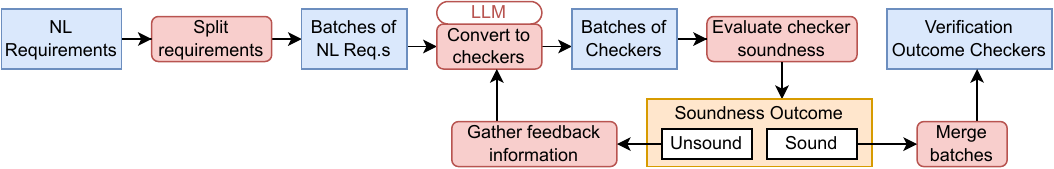}
    \caption{Overview of our divide-and-conquer approach for generating verification outcome checkers}
    \label{fig:DAC}
\end{figure*}

\autoref{fig:DAC} shows a detailed workflow of our proposed divide-and-conquer \textit{Checker generator} that takes as input a set of NL requirements and derives a checker for each requirement within that set.

\todo{MR7}
\textbf{Types of checkers.}
In this paper, we consider two types of checkers, namely declarative checkers based on SMT solvers and imperative checkers based on Python scripts.
\added{On one hand, a \textit{declarative checker} is the formalization of the input NL requirement set into an SMT-solver-friendly (static) specification.
This specification is fed to an SMT solver, which can verify its satisfiability.
Declarative checkers may also be used to check the consistency of a generated string by integrating the string value as a constraint within the checker specification.
On the other hand, an \textit{imperative} checker consists of, for each requirement, a function where the requirement is formalised using standard software language constructs. The function takes as input a string and returns a boolean value indicating whether the string satisfies the requirement formalization.
}
\todo{Add exmaple here?}

\textbf{Requirements splitting.}
As an initial step, the checker generator \textit{splits} the NL requirements into a set of \textit{requirement batches} which will be converted into checkers as a contiguous unit.
Splitting may be done arbitrarily, and will result in a batching that lies between the following extremes (which we assess in our evaluation):
% in this paper, we evaluate the performance two splitting approaches:
(1) \textit{complete split}, which yields batches containing a single requirement, and (2) \textit{no split}, which considers the entire requirement set as a single batch.

Such an input batching approach is widely used to enhance efficiency across various domains, including neural network training and testing \cite{kandel2020batchsize}, as it often helps reduce neural network's memory usage and/or training time. For our checker generation, batching reduces the number of LLM calls required to generate all checkers and may provide context when requirements are inter-related. However,
for this paper, we hypothesize that complete splitting reduces the complexity of checker generation despite an increased number of LLM calls. We will demonstrate this in \autoref{sec:evaluation}.

% batching creates a tradeoff between the number of required LLM calls (which influences cost) and the precision of (i.e. the variety of requirements addressed by) each LLM call.
% Such a tradeoff may be optimized in accordance to verification requirements

%%%%%%%%%%%%%%%%%%%%%
\textbf{Requirements-to-checker conversion.}
Each batch of NL requirements is then converted into a corresponding batch of checkers using an LLM.

For \added{generating imperative} Python checkers, we employ \textit{zero-shot} prompting, where the prompt consists of a Python function signature and a docstring comment containing the NL requirement. This approach specifies the intended function behavior, similar to how the HumanEval dataset prompts LLMs in their code generation benchmark \cite{chen2021evaluating}.

For \added{generating declarative} SMT checkers, we provide one example in the prompt consisting of a batch of NL requirements along with an example pair of NL requirements and their corresponding SMT checker \removed{output}\added{expression in string format}, and it focuses on deriving exactly one \added{SMT} \added{expression string as} checker for each requirement within the batch.
\todo{R3-5}
This example guides the LLM on the expected output format, increasing the likelihood that the generated result adheres to the desired format.

To enhance the LLM's performance, we also incorporate several best practices into the prompt design, namely  \textit{role description}, \textit{chain-of-thought prompting} \cite{wei2022chain}, and \textit{emotional stimuli} \cite{li2023large}.
% The prompt begins with a \textit{role description}, establishing the task context. To improve reasoning, we enable the \textit{zero-shot chain-of-thought} \cite{kojima2022large} by instructing the LLM to think step by step. Additionally, \textit{emotional stimuli}, shown to improve LLM performance across various tasks \cite{li2023large}, are appended at the prompt's end. The scenario generator provides the string variable's \textit{name} and \textit{constraints}.

% Since LLMs generate text-based responses, a \textit{result parser} is used to extract the values produced by the model. To facilitate straightforward extraction, the LLM is instructed to output responses in a JSON format. If the LLM fails to generate a valid JSON, it is prompted to retry until a valid response is produced. 

% The LLM's output can fall into two types: (1) a string value that \textit{may} satisfy the input constraints or (2) \textbf{UNSAT} indicating that the scenario \textit{may} be unsolvable. The correctness of LLM outputs cannot be guaranteed, hence we use a constraint solver to validate them. When the LLM generates a value for the string variable, a new  constraint is introduced in the constraint solver input requiring the string variable to equal the LLM-generated value. The value is considered valid if the constraint solver returns a \textbf{SAT} result, otherwise it is invalidated. This mechanism can also act as a filtering step, prompting the LLM to retry if the generated value is invalid. 

%%%%%%%%%%%%%%%%%%%%%
\textbf{Checker soundness evaluation.}
Once a batch of checkers is derived, it is subjected to a syntactic \textit{soundness evaluation}, which may be augmented with semantic soundness.
Note that for a given batch of checkers, the soundness of each individual checker is evaluated separately.
If all checkers within the batch are deemed to be sound, the batch is ready to be merged with other batches and returned as a contiguous set.
Otherwise, if there exist unsound checkers in a batch, feedback information is gathered from the soundness evaluation (e.g., which specific checkers are unsound), then fed back into the LLM-based requirement-to-checker converter.
Once all batches of checkers are evaluated to be sound, they are merged into a complete set of verification outcome checkers to be returned.

Since checkers are formal textual artifacts, their soundness is two-fold: syntactic and semantic.
On one hand, checkers must conform to the syntactic rules of the corresponding formal language, e.g., Python syntax for imperative checkers.
This may be evaluated efficiently by language-specific parsers.

On the other hand, ensuring \textit{semantic} soundness of checkers (i.e., whether the checker indeed checks the correct property) is complex and formalism-dependent.
For declarative checkers, this may be assessed via formal equivalence checks comparing them to ground truth implementations, if available.
When an individual NL requirement is satisfiable, semantic correctness, for both declarative and imperative checkers, may be approximated through testing, as proposed by existing research \cite{chen2021evaluating}.
In this setting, valid- and invalid-labeled data is fed into the checker and the derived checker outcome is compared to the ground-truth labeling.
% \todo{Remove this last part?}

\section{Checker-based NL Requirements Verifier}
\label{sec:loop2}

\begin{figure}
    \centering
    \includegraphics[width=0.95\linewidth]{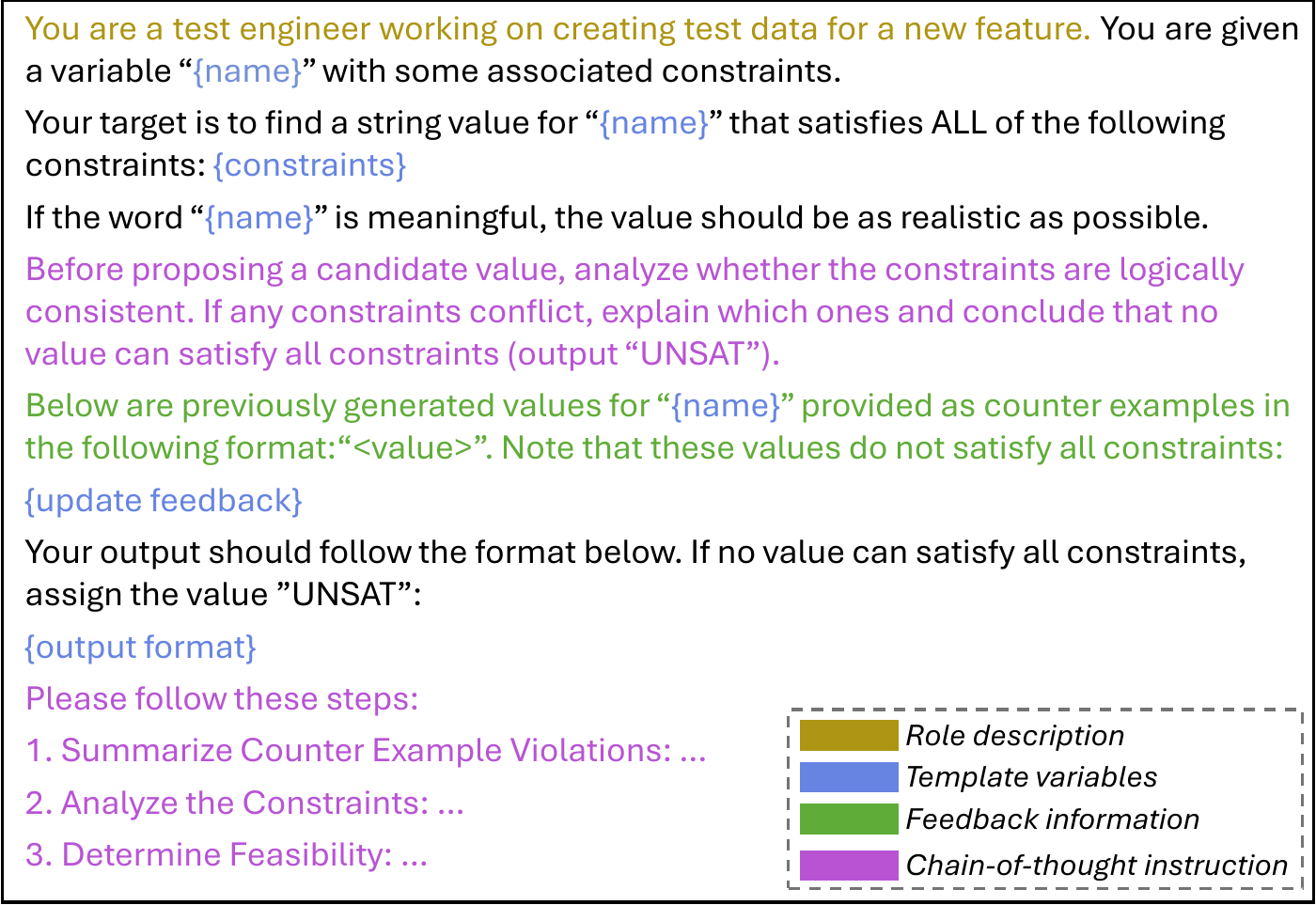}
    \caption{LVO generation with feedback prompt template excerpt}
    \label{fig:lvo-prompt}
\end{figure}

\begin{figure*}[htp]
    \centering
    \includegraphics[width=0.8\linewidth]{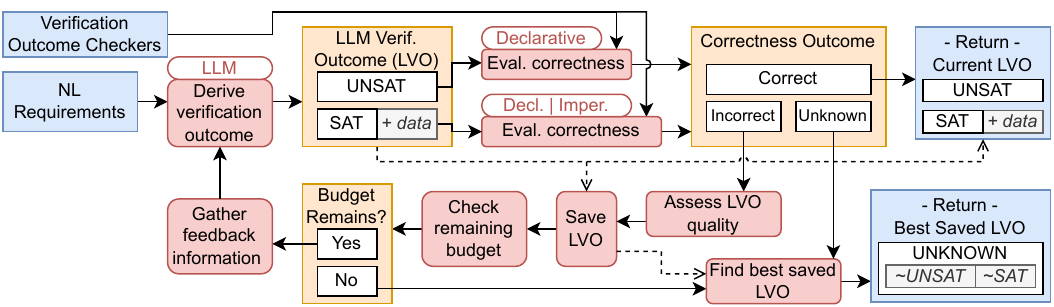}
    \caption{Overview of our checker-based NL requirements verification approach}
    \label{fig:TDG}
\end{figure*}

\autoref{fig:TDG} shows a detailed workflow of our proposed checker-based NL requirements verifier component, which takes as input the original \textit{NL requirements} and the \textit{verification outcome checkers} derived in accordance to \autoref{fig:DAC} and returns a verification outcome for the input requirements.
As an initial step, we leverage an LLM to derive an \textit{LLM-derived verification outcome} (LVO) for the input NL requirements.
For this purpose, we adopt \textit{zero-shot} prompting by describing the requirement verification in the prompt. We then ask the LLM to output string data that satisfies all the NL requirements if it is possible to do so.
Otherwise, we ask the LLM to output \textit{UNSAT}.
We follow the similar prompting best practices as described earlier in \autoref{sec:loop1}. \added{\autoref{fig:lvo-prompt} shows an excerpt of the prompt template used for generating LVOs with feedback.}

% \begin{tcolorbox}[boxsep=-1mm]
%   \added{
% You are a test engineer working on creating test data for a new feature. You are given a variable "\{name\}" with some associated constraints.

% Your target is to find a string value for "\{name\}" that satisfies ALL of the following constraints:
% \{constraints\}
% If the word "\{name\}" is meaningful, the value should be as realistic as possible.

% Before proposing a candidate value, analyze whether the constraints are logically consistent. If any constraints conflict, explain which ones and conclude that no value can satisfy all constraints (output "UNSAT").

% Below are previously generated values for "\{name\}" provided as counter examples in the following format: "{\textless}value{\textgreater}". Note that these values do not satisfy all constraints:

% \{update\_feedback\}

% Your output should follow the format below. If no value can satisfy all constraints, assign the value "UNSAT":

% \{output\_format\}
%   }
% \end{tcolorbox}
% \captionof{figure}{LVO generation prompt template with feedback.}
% \label{fig:lvo-prompt}
% \vspace{0.5em}

% \textbf{LLM-based verification outcome derivation.}

\textbf{Correctness evaluation.}
Once the LVO is derived, we \textit{evaluate its correctness} by leveraging the checkers given as input.
When the LLM claims that the input requirements are unsatisfiable, our approach relies on declarative checkers (e.g., based on SMT solving) to assess its correctness.
When the LLM provides a proof-of-consistency string as output, both declarative and imperative (e.g., a Python script) checkers may be leveraged for correctness evaluation.
This separation is due to the theoretical limits of imperative checkers: they rely on queries and operations over data to provide a correctness outcome, which is not possible if no data is provided by the LLM (as in the \textit{UNSAT} case).

If the checker confirms the correctness of the LVO, the latter (i.e., either \textit{UNSAT} or \textit{SAT+data}) is returned by the process.
Otherwise, the checker may return an \textit{incorrect} or an \textit{unknown} outcome.
An \textit{incorrect} outcome triggers a feedback loop, where the \textit{quality} of the current LVO is assessed \added{\textit{respective to the configured level of feedback information} (e.g., the specific requirements that were violated)}, then the LVO \added{and its feedback} is \textit{saved} for future use.
\todo{MR5}
The quality of an LVO is determined as a side-effect of its correctness evaluation.
In the case where the LVO is \textit{SAT+data} and the checker returns \textit{UNSAT}, we associate the quality of the LVO to the ratio of input requirements that the \textit{data} satisfies.
If the LVO is \textit{UNSAT} and the checker returns \textit{SAT}, then the quality of the \textit{UNSAT} LVO is set to a fixed value.
Note that the SMT checker may also produce data for the input requirements. We do not wish to return the checker-generated data directly since data generated by SMT checkers often lacks realism, and the LLM-generated checkers themselves may also be incorrect.

Declarative checkers may also return an \textit{unknown} correctness outcome due to limitations of the checker (e.g., the SMT process). Thus, the feedback loop is not initiated.
Instead, \textit{the best saved LVO} is determined, as detailed below, and returned.

\textbf{Budget check.}
Once the current LVO is saved, we \textit{check the remaining budget} (e.g., runtime, number of LLM calls, or number of tokens).
If a budget remains, we continue the feedback loop by \removed{\textit{gathering different levels of feedback information} from the correctness evaluation (e.g., counterexamples and the specific requirements that were violated by them)} \added{optionally incorporating \textit{saved LVOs and/or their feedback} into the \textit{update\_feedback} field of the subsequent LVO generation prompt}.
\added{Section \ref{sec:evaluation} discusses the different levels of feedback information we used in our experiment.}
\todo{MR5}

If, at the current stage, no budget remains, we determine the best LVO among all saved LVOs (i.e., the closest-to-sound) and return an \textit{UNKNOWN} verification outcome. \added{In practice, users may proceed with such fallback output even when full correctness was not established. This behavior is particularly useful in scenarios such as interactive workflows, where quick iterations may be prioritized over absolute certainty.}
\todo{MR8}

We determine the best saved LVO as follows:
(1) If there are more saved \textit{UNSAT} LVOs than there are saved \textit{SAT+data} LVOs, then the best saved LVO (the most likely NL requirements verification outcome) is \textit{UNSAT} ($\sim$\textit{UNSAT}).
(2) Otherwise, the best saved LVO is the \textit{SAT+data} outcome with the highest ratio of input NL requirements satisfied.
In this case, the NL requirements' LVO is most likely \textit{SAT} ($\sim$\textit{SAT}).

\added{\textbf{Failure mode.}
A critical failure mode occurs when \textit{both the LLM output and its checker are flawed but agree}, leading to incorrect results. While feedback helps, it cannot fully eliminate the risk. Future work should explore safeguards like redundant or cross-validated checkers.}
\todo{R2-3}
% In this case, we then return an \textit{unknown} verification outcome, which we augment with the most-likely LVO.

\section{Evaluation}
\label{sec:evaluation}
We systematically evaluate our proposed LLM-based approach for requirement satisfiability verification.
% and via data generation.
% \todo{MR4}
% \removed{We first assess the \textit{checker generator} (RQ1) and \textit{checker-based verifier} (RQ2) independently to identify the most promising configurations for each step. Subsequently, we evaluate the quality of LVOs from the end-to-end system by integrating components with the best-performing configurations identified previously (RQ3).}
\added{
The proposed framework consists of two main components: the \emph{checker generator} and the \emph{checker-based verifier}. To assess the performance of each component individually, RQ1 evaluates the checker generator using different LLMs, while RQ2 measures the LLM verifier’s theoretical performance in an experimental setting where ground-truth checkers are available. However, these two research questions focus on isolated components rather than the overall system. Therefore, RQ3 then evaluates how well the approach performs in an end-to-end, realistic deployment setting (i.e., without ground truth checkers), where LLMs are used to (1) generate the checkers, and to (2) use them to produce LVOs based solely on NL input.
}
% of  setup for different string data generation approaches based on either constraint solvers or LLMs. Specifically, we compare three constraint solvers and three LLMs. We evaluate the their performance based on their success rate and the recall to identify unsolvable cases. Finally, we discuss the findings and implications of these experiments.
Specifically, we address the following research questions: 
\begin{enumerate}
    % \item What is the success rate of LLMs and constraint solvers in generating string values satisfying all constraints
    \item [RQ1] What is the soundness of declarative and imperative checkers generated by LLMs?
    % \item [RQ2] How do the feedback form and checker type impact LLMs' performance in generation and satisfiability checking?
    \item [RQ2] How do the \textit{feedback mechanism} and \textit{checker type} impact the quality of LVOs from the LLMs?
    % generation and satisfiability checking?
    \item [RQ3] \added{What is the end-to-end performance of the framework when LLMs generate both checkers and LVOs}?

    % \item How realistic are generated strings?
\end{enumerate}

\subsection{Evaluation setup}

\paragraph{LLMs}
We compare four popular LLMs that differ in size and accessibility. \llm{GPT-4o-mini} and \llm{GPT-4o} \cite{hurst2024gpt} are powerful LLMs recently developed by OpenAI. Compared to their predecessors, they demonstrate superior performance as well as improved speed and cost-efficiency across a range of language understanding and reasoning benchmarks. Additionally, these models exhibit strong code-generation capabilities and have been integrated into various software development tools, including GitHub Copilot \cite{2025copilot}. 

% A recent study \cite{babikian2025exploring} has also reported their impressive ability to generate consistent text data aligned with natural language requirements.

Despite their impressive capabilities, GPTs are closed-source and accessible only through APIs, which may raise concerns regarding data security and privacy. To address these potential issues, we include two open-source LLMs in our evaluation. \llm{DeepSeek V3} \cite{liu2024deepseek}, a mixture-of-experts LLM, achieves performance comparable to GPT-4o while substantially reducing training costs.  \llm{Llama3.1 8b} \cite{grattafiori2024llama} is a smaller-scale model offering a balanced trade-off between performance and computational efficiency. It has been shown to achieve performance comparable to larger-scale LLMs in various tasks \cite{grattafiori2024llama} while remaining deployable on consumer-grade GPUs.

\paragraph{Checkers}
In this experiment, we explore two types of checkers for string data validation and propose a hybrid approach that leverages the strengths of both.

\textit{(Declarative)} SMT solvers are often used for handling constraint satisfaction problems involving strings. They can verify whether a given string complies with a set of \textit{formal} constraints and detect the unsatisfiability of these constraints.
In this study, we adapt the \llm{CVC5} solver to validate strings against formal constraints.
Nevertheless, our approach is independent of the solver used, allowing other SMT solvers for strings, such as \llm{Z3} and \llm{Z3str}, to be used as well.
% Existing research indicates that \llm{CVC5} outperforms other SMT solvers, such as \llm{Z3} and \llm{Z3str} \cite{babikian2025exploring}.
However, the declarative nature of SMT solvers makes them challenging for handling complex constraints involving strings. Our preliminary experiments indicate that these solvers frequently time out on recursive constraints (e.g., placement of parentheses in mathematical formulas) while exhibiting inherent limitations with universal quantifiers \cite{Cvc5SrcTheory}. 
% \cite{Cvc5SrcTheory}
% \cite{TODO}
% \todo{To discuss: Cite the GitHub issue?}

\textit{(Imperative)} These limitations can effectively be mitigated by imperatively checking strings against constraints.  In our experiments, we select \textbf{Python} as the language for implementing imperative checkers, primarily due to the exceptional code-generation capabilities of most LLMs using Python. 
% \added{Therefore, the limitations of declarative checkers can effectively be mitigated by imperatively checking strings against constraints.}
Nevertheless, imperative checkers have an intrinsic limitation: they can only confirm that a specific string conforms to requirements while they are unable to determine if the requirements themselves are unsatisfiable.

\added{For our experiment, we ask an LLM to generate declarative checkers as an SMT string expression, which we use with Python's Z3 API to verify against our LVOs. The imperative checkers are generated as Python function implementations, which can be parsed natively to Python executables. An example of checker generation is shown in \autoref{fig:example-requirement}.}

\begin{tcolorbox}[boxsep=-1mm]
\textbf{Natural language requirement.}

The system shall require an email address to contain
exactly one “@” character.
% \newline
\newline
\textbf{Declarative checker.}

(and (str.contains s “@”) (not (str.in.re s  (re.++ (re.* re.allchar) (str.to.re “@”) (re.* re.allchar) (str.to.re “@”) (re.* re.allchar)))))
% \newline
\newline
\textbf{Imperative checker.}

def this\_constraint(s: str) -\textgreater \ bool:

\ \ \ \ return s.count(“@”) == 1
\end{tcolorbox}
\noindent\begin{minipage}{\linewidth}
\captionof{figure}{An example requirement and its according checkers.}
\label{fig:example-requirement}
\end{minipage}
% \vspace{0.5em}

\paragraph{Dataset}

% \textbf{Data collection.}
To evaluate the LVO quality of our requirements verification approach, we use requirements over twelve categories of string variables drawn from common programming exercises in technical interviews, namely \textit{Email}, \textit{Name}, \textit{Password}, \textit{Url}, \textit{Date}, \textit{IBAN number}, \textit{ISBN number}, \textit{Arithmetic expression}, \textit{Palindrome}, \textit{Parentheses sequence}, \textit{DNA sequence}, and \textit{File path}. 
\todo{MR3}
\added{While these are common programming exercises, prior studies have shown that they remain challenging for LLMs \cite{babikian2025exploring}.
Notably, the requirements in our case studies correspond to production rules derived from regular and context-free grammars, which require computational models up to pushdown automata for resolution.}
% \todoab{Gunter, Daniel, please improve the formalization of this sentence.}

% We integrate six categories used for evaluation in existing research \cite{babikian2025exploring}, namely \textit{Email}, \textit{Name}, \textit{Password}, \textit{Url}, \textit{Date}, \textit{IBAN number}.
% We extend them with six additional categories, namely \textit{ISBN number}, \textit{Arithmetic expression}, \textit{Palindrome}, \textit{Parentheses sequence}, \textit{DNA sequence}, and \textit{File path}.
% detailed. \autoref{tab:dataset} shows the newly added categories and the number of requirements associated with each category.
From these requirements, we derive a total of 340 NL requirement sets (which include both satisfiable and unsatisfiable cases) by considering all possible combinations of the requirements and their negations using \textit{equivalence partitioning} \cite{burnstein2006practical}. To ensure conditions are precise, we manually formalize these requirements.

As an example, let us consider the following requirement from the \textit{ISBN number} category " ``\textit{The system shall require an ISBN number to contain exactly 10 characters, excluding hyphens}'', which is itself satisfiable.
When fed into the \llm{GPT-4o} LLM for verification, the ''1-234-56789-X`` string is returned as a proof of satisfiability.
Satisfiability is also confirmed by the \llm{CVC5} SMT solver, which yields ''0000000000`` (a much less realistic value).

% \textbf{Ground truth labeling.}
To assess the quality of verification outputs, we first derive a ground truth satisfiability label for each requirement set by manually translating the requirements to formal logical constraints (using the SMT-LIB2 syntax) and feeding them to the \llm{CVC5} solver (with 5-second timeout).
Among the 340 requirement sets within the dataset, 230 returned \textit{SAT}, 54 returned \textit{UNSAT}, and 56 returned \textit{UNKNOWN} (which, upon manual inspection, yielded 53 \textit{SAT} cases and 3 \textit{UNSAT} cases).
Therefore, in our final labeling, we have 283 \textit{SAT} cases and 57 \textit{UNSAT} cases.

\paragraph{Implementation details}
% \todo[inline]{TODO}
% \todo[inline]{Temperature, API usages, budget in terms of number of LLM calls, we map most-likely SAT and UNSAT into actual SAT and UNSAT}

To minimize variations caused by LLM generation, we set the temperature to 0 for open-source models and to 0.01 for GPT models. When controlled variation is needed, such as retrying without feedback, we use a low temperature of 0.1. We set the maximum retry budget to 2 LLM calls for the check generator and to 5 for the checker-based verifier. For evaluation purposes, we map the most likely \textit{UNSAT} ($\sim$\textit{UNSAT}) and \textit{SAT} ($\sim$\textit{SAT}) outcomes directly to \textit{UNSAT} and \textit{SAT}, respectively. The anonymized artifact accompanying this paper is available online~\cite{artifact_re2025}.

\subsection{RQ1: Checker generation}
\paragraph{Setup}
Here, we investigate the soundness of LLMs to generate declarative (\textbf{SMT}) and imperative (\textbf{Python}) checkers from NL requirements under different settings. For both types of checkers, we provide feedback on their \textit{syntactic soundness} by the corresponding parser during checker generation.

\textbf{Compared approaches.} 
% We compare two methods for checker generation given a set of constraints: (1) The \batch{} method provides the complete set of constraints simultaneously to the LLM. Here, we group all constraints and their negations into two batches and supply these batches to the LLM to generate \textit{one checker per constraint}. (2) The \ind{} method follows a divide-and-conquer strategy. In this approach, given our assumption that the constraints on a string value are connected by conjunction, each constraint is presented to the LLM independently. The final checker is then constructed by logically combining individual checkers through conjunction.
We compare two batching methods for each requirement set, in accordance with \autoref{sec:loop1}: (1) the \batch{} method yields two batches: one for all positive requirements, one for all negated requirements;
(2) The \ind{} method creates one independent batch for each requirement.

\textbf{Metrics.}
We evaluate the quality of $n$ generated checkers from both syntactic and semantic perspectives. 
% For each aspect, we measure performance using \emph{accuracy}, defined as the ratio of the number of valid results to the total number of samples.

\textit{For syntactic soundness}, we rely on Python and SMT parsers to verify the syntactic correctness of generated checkers. Formally, let $\set{V}$ be the set of generated checkers and $\texttt{P}: \set{V} \to \{0,1\}$ represent a parser function that returns 1 if the checker is syntactically valid and 0 otherwise. We define \textbf{syntax accuracy} as the ratio of syntactically valid checkers to the total number of checkers generated: $\frac{\sum_{v \in \set{V}} \texttt{P}(v)}{n} $.
Such 
% a level of abstraction 
a metric is adequate to evaluate the usability of generated checkers (compared to, e.g., a finer-grain syntax accuracy metric that counts the syntax errors) since checkers must be syntactically error-free to be applicable in our use case.
% \todo{R3-11}
% We define the \textbf{syntax accuracy} as the proportion of syntactically correct checkers to the total number of checkers generated.

% Syntax accuracy alone does not guarantee that the generated checker accurately reflects the intended natural language requirements.
We also define \textit{the proportion of \textbf{SMT} checkers whose equivalence with ground truth checkers has been formally proven} as \textbf{formal accuracy}. Formally, for a set of checkers $\set{V}$, the formal accuracy is defined as $\frac{|\{v|v\in\set{V} \land v \equiv v_{gt}\}|}{n}$. This equivalence check is converted into an SMT formulation and subsequently verified using the CVC5 solver. Checkers are considered formally correct only if equivalence is \textit{explicitly} proven; all solver results that yield \textit{timeout} or \textit{unknown} outcomes are disregarded. Formal accuracy is the most precise metric for evaluating correctness; however, it can only be computed using declarative checkers, as it requires the CVC5 solver to formally verify the equivalence of two SMT expressions. To the best of our knowledge, no existing method allows for the formal equivalence checking of Python functions.
% \todo{R2-6}

\textit{To evaluate semantic soundness \added{across both declarative and imperative checkers}}, \added{w}e adopt a testing-based approach. Specifically, for each requirement, we manually construct five string samples that satisfy the requirement and five that violate it, resulting in a total of 10 testing samples per requirement. These samples are then validated using the generated checkers and compared against the ground truth labels. A checker is considered semantically correct if it produces identical outputs to the ground truth checker for \emph{all} associated samples. We refer to the proportion of checkers meeting this criterion as \textbf{testing accuracy}. Formally, let $\set{S}_v$ be the set of sample string data associated with the original requirement of checker $v$, let $v_{gt}$ denote the ground truth checker corresponding to $v$, and let $\texttt{E}(v, s)$ represent the execution output of checker $v$ on sample $s$. Testing accuracy is defined as: $\frac{|\{v|v\in\set{V} \land \forall s\in\set{S}_v: \texttt{E}(v,s)=\texttt{E}(v_{gt},s)\}|}{n}$.

Testing accuracy provides an \emph{over-approximation} of checkers' semantic correctness, as it relies on a non-exhaustive set of test samples. For \textbf{SMT} checkers, however, it is possible to \textit{formally prove} semantic correctness by verifying their equivalence with the ground truth checkers.
% \removed{We define the proportion of \textbf{SMT} checkers whose equivalence with ground truth checkers has been formally proven as \textbf{formal accuracy}. Formally, for a set of checkers $\set{V}$, the formal accuracy is defined as $\frac{|\{v|v\in\set{V} \land v \equiv v_{gt}\}|}{n}$. This equivalence check is converted into an SMT formulation and subsequently verified using the CVC5 solver. Checkers are considered formally correct only if equivalence is \textit{explicitly} proven; all solver results that yield \textit{timeout} or \textit{unknown} outcomes are disregarded.}

\begin{table}[tb]
    \centering
    \caption{Accuracy of \textbf{Python} checker generation (in \%)}
    \begin{tabular}{c|c|c|c}
    LLM & Approach & Syntax Acc & Testing Acc \\
    \hline
    \multirow{2}{*}{\llm{Llama3.1-8b}} & \batch{} & 100 & 100\\
    & \ind{} & 100 & 98 \\
    \hline
    \multirowcell{3}{\llm{GPT-4o-mini} \\ \llm{GPT-4o} \\ \llm{DeepSeek-V3}} & \multirowcell{3}{\batch{}\\ \ind{}} & \multirowcell{3}{100\\ 100} & \multirowcell{3}{100\\ 100}\\
     & & & \\
    & & & \\
    % & & & \\
    % & \ind{} & 100 & 100\\
    \hline
    \end{tabular}
    \label{tab:python_checker}
\end{table}

\begin{figure}[tb]
    \centering
    \includegraphics[width=0.8\linewidth]{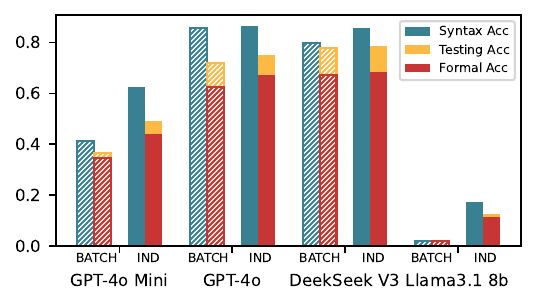}
    \caption{Accuracy of generated \textbf{SMT} constraints (in \%)}
    \label{fig:smt_checker}
\end{figure}

\paragraph{Results}
\autoref{tab:python_checker} and \autoref{fig:smt_checker} present the soundness of \textbf{Python} and \textbf{SMT} checkers generated by LLMs, respectively. For Python, all tested LLMs exhibit exceptional capability, achieving perfect syntax accuracy across both \batch{} and \ind{}. Moreover, three of the evaluated models (\llm{DeepSeek-V3}, \llm{GPT-4o}, and \llm{GPT-4o-mini}) maintain perfect testing accuracy (100\%), with only \llm{Llama3.1-8b} slightly lower at approximately $98\%$ in the \ind{} setting. These results strongly suggest that Python is a highly suitable language for checker generation using LLMs, leveraging their strengths in Python code generation tasks. Indeed, prior research has already demonstrated LLMs' capability on challenging code generation benchmarks such as HumanEval \cite{liu2024deepseek}; generating Python-based checkers for string validation is then a comparatively less challenging task. However, it is important to note that Python checkers are not capable of validating unsatisfiability.

LLMs' capability is particularly important when generating \textbf{SMT} checkers. As shown in \autoref{fig:smt_checker}, smaller-scale LLMs (\llm{GPT-4o-mini} and \llm{Llama3.1-8b}) achieve notably lower quality compared to larger models (\llm{GPT-4o} and \llm{DeepSeek-V3}). Specifically, outputs from \llm{Llama3.1-8b} exhibit significantly lower quality, achieving less than 20\% syntax and testing accuracy in \ind{}. This clearly suggests that \llm{Llama3.1-8b}'s smaller scale limits its ability to effectively generate SMT checkers. Conversely, \llm{DeepSeek-V3} and \llm{GPT-4o} achieve substantially higher accuracy, both with over 85\% syntax accuracy and over 75\% testing accuracy in \ind{}. Interestingly, formal accuracy is closely aligned with testing accuracy, suggesting that most of the checkers passing the sample-based tests can be formally proven correct.
When comparing syntax correctness to semantic correctness, the ratio between syntax accuracy and semantic accuracy exceeds 65\% in all evaluated cases and surpasses 70\% in six out of eight cases. This observation suggests that while syntax accuracy is already relatively high, further enhancing syntax correctness could still positively influence semantic soundness.

Comparing \batch{} and \ind{} generation approaches shows that treating each requirement independently generally benefits the LLMs, with \ind{} consistently outperforming \batch{} across all tested LLMs. The accuracy gain from independent handling is particularly pronounced for smaller LLMs. For instance, \llm{GPT-4o-mini} shows an over 10\% increase in accuracy when using \ind{}, compared to only around 3\% improvement for \llm{GPT-4o}. This indicates that larger LLMs can effectively handle multiple requirements simultaneously, making the simplified \ind{} scenario less beneficial to them.

% \begin{enumerate}
%     \item Python in general achieves perfect results
%     \item For SMT, LLM's capability matters
%     \item Idenpendent generation is in general better than batch, especially for smaller LLMs
%     \item Most SmT constraints can be formally verified.
    
% \end{enumerate}

\begin{tcolorbox}[boxsep=-1mm]
\textbf{Answer to RQ1.}
Three LLMs achieve perfect accuracy (100\% testing accuracy) in generating Python checkers, with \llm{Llama3.1-8b} achieving near-perfect accuracy ($\geq$98\%) across both \batch{} and \ind{} settings. For SMT checkers, LLM size significantly influences the soundness, as larger LLMs (over 75\% in testing accuracy for \llm{DeekSeek-V3}) consistently outperform smaller ones (less than 20\% in testing accuracy for \llm{Llama3.1-8b}). Moreover, treating each requirement independently (\ind{}) consistently helps improve both syntactic and semantic soundness compared with batched input (\batch{}).
\end{tcolorbox}

\subsection{RQ2: Checker-based verification}
% \subsection{RQ2: Checker-based verification}

\paragraph{Setup}
% Existing work suggests that using human-created SMT checkers to validate test data generated by LLMs and retrying generation when inconsistencies with the original requirement occur may significantly improve the quality of generated texts \cite{babikian2025exploring}. 
Existing studies suggest that applying an explicit validation stage for LLM generation may significantly improve the LLM's performance in various tasks \cite{babikian2025exploring, li2024guiding,first2023baldur,kirchner2024prover}.
Building upon this insight, this research question further explores how different feedback mechanisms and checkers influence the quality of LVO produced by LLMs in the context of requirement verification. Specifically, this RQ focuses on the theoretical performance of checker-based verification when a ground truth checker is present. In RQ3, the best theoretical settings are then evaluated in the end-to-end setting to understand the overall effectiveness in a realistic use case. 

\textbf{Compared approaches.} 
We compare the direct LLM generation approach with various checkers and feedback methods. The \textbf{direct} approach validates the satisfiability of constraints with a single call to the LLM.
\todo{R3-13}
\added{In this approach, the NL requirements are fed into the LLM, and the derived LVO is returned directly (with neither a checker-based evaluation nor a feedback loop). In this case, a checker-based evaluation is irrelevant since the derived feedback cannot be leveraged.}

For feedback approaches, we assess both imperative checkers in \textbf{Python} and declarative checkers using \textbf{SMT}. Moreover, we propose a \textbf{Hybrid} approach, which uses \textbf{Python} to validate generated strings and \textbf{SMT} to verify unsatisfiability decisions.

For each checker type, we examine three distinct feedback mechanisms: (1) \textbf{V} (validation), which follows a trial-and-error approach,
% from an existing research \cite{babikian2025exploring}, 
where checkers provide no explicit feedback but ask the LLM to retry upon incorrect outcomes; (2) \textbf{VF} (validation with feedback), which explicitly incorporates \added{incorrect LVOs from previous loop(s)} as counterexamples \added{within the generation prompt}; and (3) \textbf{VFE} (validation with feedback and explanation), which further enriches feedback by 
% \removed{including detailed explanations of failed requirements alongside each counterexample} 
\added{specifying which of the natural language requirements did each counterexample fail, using validation results from each requirement's corresponding checker}.
\todo{MR5 / R3-12}

\textbf{Metrics.}
We assess LVOs from different approaches based on the generation quality of the test data for \textit{SAT} outcome and the quality of \textit{UNSAT} decisions.

The Generation Success Rate (\textbf{GSR}) measures the percentage of requirement sets successfully verified by generating appropriate test data. Formally, let $\set{D}$ be the dataset of requirements, $\set{D}_{\texttt{sat}} \subseteq \set{D}$ represent the subset of satisfiable requirements, $v^{d}_{gt}$ denote the ground truth checker associated with each set of requirements $d \in \set{D}$, and $s_d$ represent the string generated by an approach for a requirement set $d$. Furthermore, let $\texttt{E}(v, s) \in \mathbb{B}$ indicate the validation result of checker $v$ on output $s$. The \textbf{GSR} is defined as: $GSR=\frac{|\{d|d\in \set{D}_{sat}, \texttt{E}(v^d_{gt}, s_d)\}|}{|\set{D}_{\texttt{sat}}|}$.

To evaluate the effectiveness of each approach in detecting unsatisfiable requirement sets, we compute standard precision (\textbf{P}), recall (\textbf{R}), and \textbf{F1}-score metrics using the unsatisfiable subset of the dataset, $\set{D}_{\texttt{unsat}}$. Let $\set{S}_{\texttt{unsat}}$ denote the set of outputs identified as unsatisfiable by an approach. These metrics are defined as: $P=\frac{|\set{D}_{unsat} \cap \set{S}_{unsat}|}{|\set{S}_{unsat}|}$, $R=\frac{|\set{D}_{unsat} \cap \set{S}_{unsat}|}{|\set{D}_{unsat}|}$, and $F1=\frac{2 \times P \times R}{P + R}$. 

 \begin{table*}[tb]
    \centering
    % \caption{Performance of compared approaches on string test data generation (in \%)}
    \caption{Quality of LVOs from compared approaches on NL requirements verification with ground truth checkers (in \%)}
    \label{tab:rq1}
    \scalebox{0.9}{
    \begin{tabular}{|c|c|cccc|cccc|cccc|cccc|}
        \hline
        & & \multicolumn{4}{c|}{GPT-4o-mini} & \multicolumn{4}{c|}{GPT-4o}  & \multicolumn{4}{c|}{DeekSeek-V3} & \multicolumn{4}{c|}{Llama3.1-8b} \\
        \hline
        checker & Methods & \gsr & \precision & \recall & \fscore & \gsr & \precision & \recall & \fscore & \gsr & \precision & \recall & \fscore & \gsr & \precision & \recall & \fscore \\
        \hhline{|==================|}
         & Direct & 66.43 & 58.33 & 61.40 & 59.83 & 78.45 & 90.32 & 49.12 & 63.64 & 72.08 & 63.27 & 54.39 & 58.49 & 25.80 & 35.29 & 33.33 & 34.29\\
        \hline
        \multirow{3}{*}{\python} & +V & 73.50 & 47.37 & \underline{63.16} & 54.14 & \underline{91.87} & 87.50 & 49.12 & 62.92 & \underline{89.75} & \underline{70.45} & 54.39 & 61.39 & 38.16 & 36.00 & 31.58 & 33.64\\
        & +VF & 71.38 & 50.77 & 57.89 & 54.10 & 90.81 & \textbf{\underline{96.67}} & \underline{52.63} & \underline{68.16} & 82.33 & 66.00 & \underline{57.89} & \underline{61.68} & 38.87 & 34.15 & 26.32 & 29.72\\
        & +VFE & \underline{74.56} & \underline{52.31} & 59.65 & \underline{55.74} & 91.52 & 90.62 & 50.88 & 65.17 & 84.81 & 62.00 & 54.39 & 57.94 & \underline{45.94} & \underline{46.67} & \underline{36.84} & \underline{41.18}\\
        \hline
        \multirow{3}{*}{\solver} & +V & 74.20 & 60.56 & 75.44 & 67.19 & 86.93 & 69.05 & 50.88 & 58.59 & 86.93 & 70.00 & 61.40 & 65.42 & \underline{55.48} & 70.91 & 68.42 & 69.64\\
        & +VF & \underline{75.97} & 65.82 & \underline{91.23} & 76.47 & 87.28 & \underline{82.26} & 89.47 & 85.71 & 84.10 & \underline{80.60} & \textbf{\underline{94.74}} & \underline{87.10} & 48.76 & \underline{71.43} & \textbf{\underline{98.25}} & \underline{82.72}\\
        & +VFE & 74.56 & \underline{}{69.33} & \underline{91.23} & \underline{78.79} & \underline{89.75} & 81.54 & \textbf{\underline{92.98}} & \underline{86.89} & \underline{89.75} & 79.41 & \textbf{\underline{94.74}} & 86.40 & 46.64 & 68.42 & 91.23 & 78.20\\
        \hline
        \multirow{3}{*}{\hybrid} & +V & 77.39 & 63.08 & 71.93 & 67.21 & \textbf{\underline{92.23}} & 93.55 & 50.88 & 65.91 & \textbf{\underline{91.87}} & 85.37 & 61.40 & 71.43 & 53.36 & 77.50 & 56.14 & 65.11\\
        & +VF & \textbf{\underline{78.09}} & 72.00 & \textbf{\underline{94.74}} & 81.82 & 90.11 & 96.15 & 87.72 & 91.74 & 90.81 & 87.10 & \textbf{\underline{94.74}} & 90.76 & 51.94 & \textbf{\underline{80.95}} & 91.23 & 85.78\\
        & +VFE & 77.74 & \textbf{\underline{74.29}} & 92.98 & \textbf{\underline{82.59}} & \textbf{\underline{92.23}} & \underline{96.36} & \textbf{\underline{92.98}} & \textbf{\underline{94.64}} & \textbf{\underline{91.87}} & \textbf{\underline{89.83}} & 96.49 & \textbf{\underline{93.04}} & \textbf{\underline{57.24}} & 80.60 & \underline{94.74} & \textbf{\underline{87.10}}\\
        \hline
    \end{tabular}
    }
\end{table*}

\paragraph{String generation results}
The measurement results for requirements verification across different checkers, feedback methods, and LLMs are presented in \autoref{tab:rq1}. Underlined numbers indicate the best quality achieved by each checker by integrating with a specific LLM, while bold numbers highlight the top-performing results for each LLM. Results from the baseline (\textbf{direct}) approach reveal a significant difference in GSR depending on the scale of the LLM. 
% consistent with existing findings \cite{babikian2025exploring}. 
Evidently, LLMs' capacity is directly correlated with output quality, as \llm{GPT-4o} achieves the highest \textbf{direct} GSR at 78.45\%, whereas smaller-scale LLMs, such as \llm{Llama3.1-8b}, notably lag at 25.80\%. Nevertheless, even the highest GSR from \textbf{direct} remains relatively low, highlighting the importance of additional validation techniques to enhance output quality.

Introducing checkers (\textbf{+V}) consistently enhances GSR across all evaluated LLMs compared to the \textbf{direct} approach. Notably, the GSR for \llm{GPT-4o} increases from 78.45\% to 91.87\% when using the \textbf{Python} checker. Smaller models exhibit even greater relative improvement; for example, the GSR of \llm{Llama3.1-8b} more than doubles from 25.80\% to 55.48\% with validation by SMT checkers. These results show the effectiveness of checkers in enhancing the capability of smaller LLMs to produce valid string outputs.

Interestingly, incorporating feedback with counterexamples (\textbf{+VF}) does not consistently enhance GSR and, in some instances, slightly decreases it. For example, using the \textbf{SMT} checker, the GSR notably drops from 55.48\% to 48.76\%. This reduction could be attributed to confusion caused by counterexamples lacking explanations, making it difficult for LLMs to identify why specific inputs fail validation. This hypothesis is reinforced by the improvements when explanatory feedback (\textbf{+VFE}) is introduced, resulting in enhanced GSR in 7 out of the 12 tested scenarios compared to both \textbf{+V} and \textbf{+VF}.

When comparing the effectiveness of different checkers, \textbf{Python} generally yields improvements similar to those of \textbf{SMT} for smaller LLMs (\llm{GPT-4o-mini} and \llm{Llama3.1-8b}). We hypothesize that smaller LLMs struggle with more complex requirements where Python checkers could be more beneficial. Indeed, \textbf{Python} consistently outperforms \textbf{SMT} for larger models, particularly \llm{GPT-4o}, achieving the highest GSR of 91.87\%, compared to 89.75\% with \textbf{SMT}. Furthermore, combining \textbf{Python} and \textbf{SMT} into a \textbf{Hybrid} checker results in even better LVO quality, providing the highest GSR of 92.23\% across all LLMs and feedback configurations evaluated.

\paragraph{Requirement verification results}
When evaluating the effectiveness of the approaches in verifying the satisfiability of requirement sets, the \textbf{Python} checker alone does not consistently improve the generation F1-score, as it cannot validate the unsatisfiability (\textit{UNSAT}) decisions made by LLMs. Conversely, the \textbf{SMT} checker can effectively verify these \textit{UNSAT} decisions, yielding significant improvements over the \textbf{direct} approach. Notably, incorporating explanatory feedback (\textbf{+VFE}) enhances precision for three out of four LLMs and improves recall and F1-scores across all evaluated models. Moreover, the \textbf{Hybrid} checker uses Python's better capability to detect invalid string outputs, further improving output quality and achieving higher F1-scores compared to the \textbf{SMT} checker alone in the \textbf{+VFE} scenario. This improvement is more notable for smaller LLMs; for example, the F1-score of \llm{Llama3.1-8b} increases substantially from 34.29\% to 87.10\% using the \textbf{Hybrid+VFE} approach.

When comparing feedback methods, providing more informative feedback consistently helps LLMs identify inconsistencies within sets of requirements. Introducing validation alone (\textbf{+V}) already enhances the F1-score in identifying unsatisfiable requirement sets, improving it in 8 out of 12 cases. However, because this method does not offer explicit feedback, quality gains remain limited. Adding counterexamples (\textbf{+VF}) further enhances the output quality, yielding gains in 10 out of 12 scenarios compared to validation alone. Supplementing these counterexamples with explanations (\textbf{+VFE}) results in the highest overall LVO quality, further improving the F1-score in 8 out of 12 scenarios compared to \textbf{+VF} cases.
% counterexamples without explanations.

% 0. Larger LLMs performs in general better than smaller LLMs
% 1. All feedback approaches are better than the direct approach in terms of GSR
% 2. Python does not improve the overall F1 of unsatifiability identification, since it cannot handle the unsat cases
% 3. SMT provides a better F1-score compared with Python. However, the hybrid approach gives overall the best performance
% 4. Only adding counterexamples may hurt the performance, particularly for smaller LLMs
% 5. Adding an explanation helps with this

\begin{tcolorbox}[boxsep=-1mm]
\textbf{Answer to RQ2.}
Incorporating feedback from ground truth checkers substantially improves LVO quality from LLMs in terms of both GSR and verification F1-scores. Specifically, the \textbf{Hybrid} checker, combined with counterexamples and explanations (\textbf{+VFE}), achieves the best results, providing the highest GSR (57.24\% to 92.23\%) in three out of four LLMs and the best verification F1-scores (82.59\% to 93.04\%) across all evaluated LLMs.

\end{tcolorbox}

\subsection{RQ3: End-to-end evaluation}
\paragraph{Setup}
In this research question, we integrate the previously evaluated \textbf{checker generator} from RQ1 and \textbf{checker-based verifiers} from RQ2 to comprehensively assess the overall system.

\textbf{Compared approaches.}
We evaluate the combined system of the best configurations identified in previous research questions for the \textbf{checker generator} and the \textbf{checker-based verifier}. Specifically, for the \textbf{checker generator}, we select the checkers generated by \ind{} across all evaluated LLMs. For the \textbf{checker-based verifier}, we adopt the \textbf{Hybrid+VFE} configuration, which integrates the advantages of both declarative and imperative validation methods and provides the most detailed feedback. We assess the effectiveness of these combined configurations across all LLMs, comparing their output quality against the baseline results obtained without any checker \added{or feedback loop} (\textbf{Direct}, as presented in RQ2).

\textbf{Metrics.} We evaluate the LVO quality from the integrated system using the same metrics as defined in RQ2: \textbf{GSR}, \textbf{P}, \textbf{R}, and \textbf{F1}. Since RQ2 configurations use ground truth checkers, their corresponding metric values represent an upper bound on achievable LVO quality for the complete system. Therefore, in this RQ, we report the \emph{ratio} of the measured metrics to these upper-bound values for each metric. Specifically, given a measured metric value $m$ and its corresponding upper-bound value $\hat{m}$ (obtained using ground truth checkers), we calculate and plot the ratio $\frac{m}{\hat{m}}$ to illustrate how closely each evaluated configuration approaches the upper-bound metric values.

\todo{There's an empty area here. Check if it's still here on the final edit.}

\begin{figure*}[htp]
    \centering
    \includegraphics[width=\linewidth]{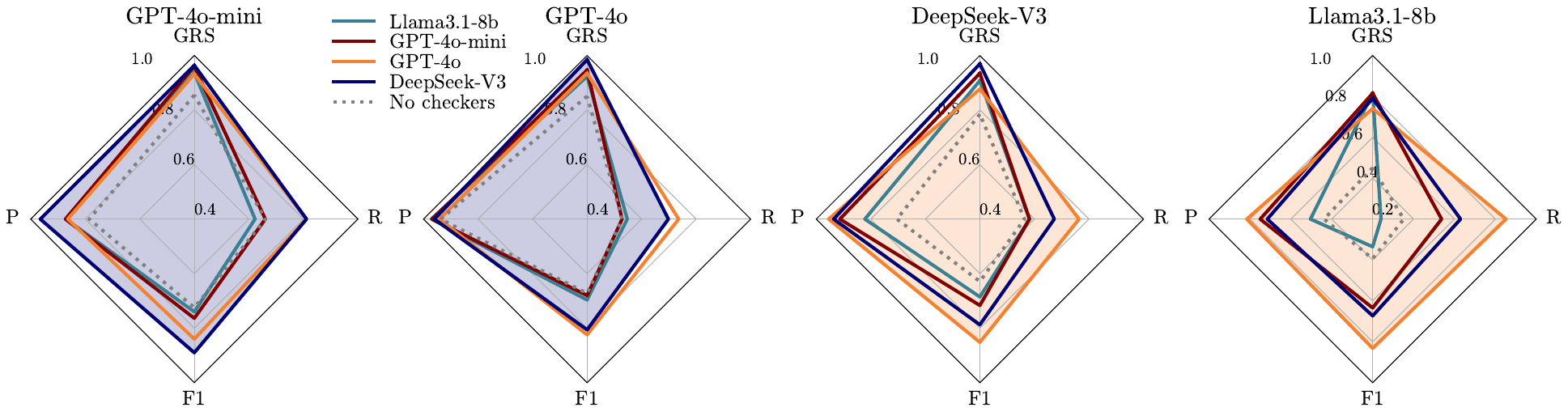}
    \caption{Metric ratio of different LLM-based verifiers using LLM-generated checkers compared with the ground truth checkers; each figure represents a different LLM-based verifier; each line represents a checker generation method; the filled color shows the best checker generation method for each verifier}
    \label{fig:rq3}
\end{figure*}

\paragraph{Results}
\autoref{fig:rq3} presents radar plots illustrating the metric ratio of the system across various LLM combinations compared against the baseline approach without validation (\textbf{Direct} from RQ2). Specifically, the center of each radar plot corresponds to a metric ratio of 0.4 for \llm{GPT-4o-mini}, \llm{GPT-4o}, and \llm{DeepSeek-V3}, whereas the center represents a ratio of 0.2 for \llm{Llama3.1-8b}.

The quality of LVOs generated using the feedback approaches significantly outperforms the configuration without checkers (\textbf{Direct}) across almost all metrics and configurations. The only exceptions occur when checkers generated by \llm{Llama3.1-8b} are used with either the \llm{GPT-4o-mini} or \llm{Llama3.1-8b} verifiers, likely due to the limited quality of these generated checkers combined with the lower capability of these smaller LLMs. Across all configurations, the \textbf{GSR} consistently exceeds the baseline without checkers and approaches the ground-truth checker, particularly when using checkers generated by \llm{DeepSeek-V3} combined with verifiers such as \llm{GPT-4o} and \llm{DeepSeek-V3}, achieving ratios over 95\%. However, verification \textbf{F1}-score remains considerably below their upper bounds, with the best case achieving around 88\% when combining the \llm{GPT-4o-mini} verifier with checkers generated by \llm{DeepSeek-V3}. This performance gap likely results from the substantially lower quality of generated SMT checkers compared to Python checkers, highlighting the need for further improvements in SMT checker generation methods.

When comparing output quality across different checkers, higher checker quality consistently correlates with improved overall output quality. Specifically, verifiers using the highest-quality checkers generated by \llm{GPT-4o} and \llm{DeepSeek-V3} generally achieve the best results. Notably, \llm{DeepSeek-V3}-generated checkers yield the highest \textbf{GSR} for three out of four verifiers, and \llm{GPT-4o}-generated checkers produce the best \textbf{F1}-scores in three out of four verifiers. Interestingly, even though checkers generated by \llm{Llama3.1-8b} have significantly lower SMT checker quality compared with other LLMs (see RQ1), they still enhance LVO quality across all metrics for the \llm{GPT-4o} and \llm{DeepSeek-V3} verifiers, and improve \textbf{GSR}, \textbf{P}, and \textbf{F1} metrics for the \llm{GPT-4o-mini} verifier compared to scenarios without checkers. This finding suggests that our feedback system generally shows robustness against variations in the quality of generated checkers.

% 1. The best approaches clearly get close to gt GSR but not for verification F1, which is justified by RQ1 that the Python performance is much better than SMT
% 2. Better validator quality means better overall performance, which means if the performance of SMT generation is improved, we will see the performance go up
% 3. Low validator quality may still help, but it has to be combined with more capable LLMs(Llama3.1 8b)

 %NOTE:  Because the generated constraints are poor for gpt-4o-mini, the checkers may misclassify unsatisfiable requirements as satisfiable

\begin{tcolorbox}[boxsep=-1mm]
\textbf{Answer to RQ3.}
Combining an LLM-based checker generator with a checker-based verifier significantly improves output quality compared to direct generation without checkers, achieving over 95\% GSR ratio in three verifiers and over 60\% F1-score ratio in all verifiers. Enhancing the quality of individual components, such as using higher-quality checkers and more capable verifier LLMs, directly translates into improved overall performance in the final LVO. This finding highlights the modularization of the system for achieving better overall outcomes.
\end{tcolorbox}

% \subsection{Discussion}

% \todo[inline]{Discuss Problematic cases?}
% \todo[inline]{TODO}

\subsection{Threats to Validity}
\label{sec:threats}
\textbf{Internal Validity:}
% For each introduced dataset example, t
The authors manually convert each requirement set into checkers and testing samples, which could introduce biases or inaccuracies. To mitigate potential issues arising from manual effort, we cross-check the created checks and samples to make sure they are consistent with each other. Moreover, the dataset is imbalanced and contains more \textit{SAT} cases than \textit{UNSAT} cases. To mitigate bias caused by this imbalance, we evaluate different metrics separately for each category. Additionally, we control for variations in LLM outputs by setting temperature parameters to zero for open-source models (e.g., \llm{Llama3.1-8b} and \llm{DeepSeek-V3}) and to a minimal value (0.01) for GPT-based models, except in cases where controlled variation is necessary (such as the \textbf{+V} scenario), in which case a temperature value of 0.1 is used.

\textbf{External Validity:}
% The requirements collected for experiments are from common programming exercises in technical interviews, representing requirements in an education setting. 
The collected requirements in the dataset may be impacted by selection bias. Therefore, our findings may differ when generalized to other contexts. 
\todo{MR3}
\added{Specifically, the original requirements may be present in the training data of LLMs, as they are derived from common programming exercises. However, the dataset includes all possible combinations of these requirements and their negations. Such unusual cases are less likely to occur in the training dataset.}
Future studies are necessary to evaluate whether our approach generalizes well to more complex, industry-scale requirements.

\textbf{Construct validity:}
We evaluate the quality of generated checkers based on semantic and syntactic correctness, consistent with general practice in code generation \cite{chen2024survey}. 
We assess the quality of LVOs on standard success rate for generation, and precision/recall/F1-score for classification.
% We assess the quality of LVOs using the appropriate metrics for both \textit{SAT} and \textit{UNSAT} cases. For \textit{SAT} cases, we evaluate the quality of the generated text data with respect to the input requirements. For \textit{UNSAT} decisions, we use standard classification metrics. 
% by directly assessing them wrt. the input requirements.

% Additonally, we mitigate threats to construct validity by distinclty considering satisfiable and unsatisfiable cases.
% , we use metrics aligned with existing studies \cite{babikian2025exploring}, which measure both data quality and the correctness of unsatisfiability decisions.
% \todo{Need a confirmation here}
\section{Related Work}

\noindent
\textbf{LLM for requirements engineering.}
% The impressive capabilities of 
LLMs are becoming increasingly popular in various stages of requirements engineering, including elicitation, modeling, and verification \cite{arora2024advancing}.

Ronanki et al. \cite{ronanki2023investigating} explore the strengths and limitations of LLMs in the elicitation process.
G{\"o}rer and Aydemir \cite{gorer2023generating} propose an LLM-assisted approach that yields interview questions for identifying new requirements.
% , LLMs assist in proposing interview questions  for identifying new requirements.
In requirement modeling, existing work has studied LLM's ability in generating domain models \cite{chen2023automated,camara2023assessment} and goal models \cite{chen2023use,de2024towards}.
% Since requirements often contain ambiguities or inconsistencies that are difficult to detect manually, 
LLMs have also been effectively used for helping identify ambiguities or inconsistent sets of requirements \cite{feng2024normative,fantechi2023inconsistency}.

In this paper, we leverage LLMs for verification of NL requirements for satisfiability.
As a key side effect, when requirement sets are satisfiable, our approach generates a conformant and realistic string instance as a proof of satisfiability, which is relevant as input data for requirements-based testing.

% \todo{not too sure about this. maybe we want to reposition this as "LLMs for satisfiability checking. TODO later}
\noindent
\textbf{LLM for test data generation.}
Due to their impressive NL  generation capability, LLMs have been increasingly applied to test data generation tasks (e.g., for testing graphical user interfaces~\cite{liu2023fill}, fuzz testing \cite{xia2024fuzz4all}, validating SMT solvers \cite{sun2023smt}) from (satisfiable) requirements given as input.
However, less focus has been placed on (1) satisfiability of requirements and on (2) assessing requirements over strings.

Existing research \cite{liu2024testing} takes as input user-written constraints for string variables and generates corresponding test cases.
% Liu et al. \cite{liu2024testing} propose a method to generate test cases for string variables from user-written constraints. However, they only focus on erroneous test case generation for bug detection without considering potential inconsistencies between constraints.
In contrast, our approach verifies the input requirements for satisfiability and outputs consistent test data when the requirements are satisfiable. 
Similarly, Babikian et al. \cite{babikian2025exploring} propose a textual test data generation approach from input NL requirements over strings.
However, their approach relies on manual formalization of NL requirements to ensure output correctness, which is a significant drawback compared to our proposed approach.
% \todo{Quick check here}
% However, their approach focuses on natural language requirements that are less strict. In contrast, this paper also explores generating test cases with the inclusion of strict formal specifications, such as constraints in the SMT-LIB2 syntax, as part of the requirements verification process. 

\textbf{Iterative generation with LLMs.}
Recent advances in LLMs have demonstrated impressive performance in various tasks. However, complex reasoning problems still pose significant challenges, motivating approaches that decompose complex tasks into iterative sub-tasks \cite{yao2023tree}. The Tree-of-Thought framework \cite{yao2023tree,long2023large} enables LLMs to approach decision-making incrementally, with the ability to trace back and revise decisions if necessary. Building upon this, the Graph-of-Thought framework \cite{besta2024graph} allows LLMs to organize their reasoning in a graph structure, enabling the parallel execution of multiple subtasks. More recently, multi-agent collaboration, employing multiple LLMs for distinct subtasks, has gained popularity \cite{hong2023metagpt,li2024more,wu2023autogen,gao2024agentscope}. Such methods leverage the strengths of different LLMs and specialized prompting techniques tailored for each subtask.

Iterative LLM-based approaches have been effectively adapted to software engineering tasks such as code generation \cite{hong2023metagpt,he2024llm}, test generation \cite{alshahwan2024automated}, and software modeling \cite{yang2024multi}. In requirement engineering, specifically, iterative methods have integrated retrieval-augmented generation for enhanced requirement traceability and compliance checking \cite{masoudifard2024leveraging}.

In this paper, we propose an iterative approach using LLMs for requirement satisfiability verification. Our approach can also be seen as a multi-agent system, wherein one agent (checker generator) generates verification checkers to validate the verification outputs produced by another agent (checker-based verifier).
% while leveraging the power of LLMs to generate realistic values for string variables.

% \noindent
% \textbf{LLM and SAT Solver for Sting constraints}

\section{Conclusion}

In this paper, we propose an approach that leverages LLMs, SMT solvers (i.e., CVC5), and imperative programming (i.e., Python) for the verification of NL requirements for satisfiability through consistent string and checker generation.
Our approach takes as input a collection of NL requirements over strings and either (1) detects unsatisfiable requirements or (2) generates a conforming string instance as a proof of satisfiability.
We experimentally show that checker generation is improved by assessing each requirement individually, while requirements verification benefits from integrating declarative and imperative checkers with detailed feedback.
These findings allow us to achieve near-ground-truth performance without manual intervention, e.g., NL requirement formalization.

As future work, we plan to extend our satisfiability verification approach for string requirements to more complex textual formalisms, such as domain-specific textual languages. 
\todo{MR2}
\added{While this work focuses on string requirements, the underlying idea of combining checker generation with data generation is expected to generalize to a broader range of requirement types, provided that a corresponding formal specification language and constraint satisfiability solving approach are available. 
For instance, our approach may be applicable to the satisfiability verification of NL requirements on the safety of autonomous vehicles, where such requirements are represented as numeric conditions over the position and assigned path of each vehicle in a traffic scenario \cite{babikianConcretizationAbstractTraffic2024}.
% For example, given the formal representation of safety criteria in autonomous vehicles, the approach can be extended for generating testing scenarios when combined with logic and numerical solvers \cite{TODO}.
We plan to explore this extension in future work.
}
Further improvements in SMT checker generation could be explored by incorporating semantic soundness feedback during the generation process.
Additionally, we plan to improve the feedback loop proposed in \autoref{fig:TDG} by integrating stochastic mutations over a collection of simultaneous LVOs, in analogy to a genetic algorithm.

\bibliographystyle{IEEEtran}
\bibliography{reference,StringConstraints}

\begin{thebibliography}{10}
\providecommand{\url}[1]{#1}
\csname url@samestyle\endcsname
\providecommand{\newblock}{\relax}
\providecommand{\bibinfo}[2]{#2}
\providecommand{\BIBentrySTDinterwordspacing}{\spaceskip=0pt\relax}
\providecommand{\BIBentryALTinterwordstretchfactor}{4}
\providecommand{\BIBentryALTinterwordspacing}{\spaceskip=\fontdimen2\font plus
\BIBentryALTinterwordstretchfactor\fontdimen3\font minus \fontdimen4\font\relax}
\providecommand{\BIBforeignlanguage}[2]{{%
\expandafter\ifx\csname l@#1\endcsname\relax
\typeout{** WARNING: IEEEtran.bst: No hyphenation pattern has been}%
\typeout{** loaded for the language `#1'. Using the pattern for}%
\typeout{** the default language instead.}%
\else
\language=\csname l@#1\endcsname
\fi
#2}}
\providecommand{\BIBdecl}{\relax}
\BIBdecl

\bibitem{feng2024normative}
N.~Feng, L.~Marsso, S.~G. Yaman, I.~Standen, Y.~Baatartogtokh, R.~Ayad, V.~O. De~Mello, B.~Townsend, H.~Bartels, A.~Cavalcanti \emph{et~al.}, ``Normative requirements operationalization with large language models,'' in \emph{RE 2024}.\hskip 1em plus 0.5em minus 0.4em\relax IEEE, 2024, pp. 129--141.

\bibitem{fantechi2023inconsistency}
A.~Fantechi, S.~Gnesi, L.~Passaro, and L.~Semini, ``Inconsistency detection in natural language requirements using {ChatGPT}: a preliminary evaluation,'' in \emph{RE 2023}.\hskip 1em plus 0.5em minus 0.4em\relax IEEE, 2023, pp. 335--340.

\bibitem{hosseiniAmbiguityGeneralityNatural2021}
M.~B. Hosseini, J.~Heaps, R.~Slavin, J.~Niu, and T.~Breaux, ``Ambiguity and {{Generality}} in {{Natural Language Privacy Policies}},'' in \emph{2021 {{IEEE}} 29th {{International Requirements Engineering Conference}} ({{RE}})}, 2021, pp. 70--81.

\bibitem{giannakopoulouGenerationFormalRequirements2020}
D.~Giannakopoulou, T.~Pressburger, A.~Mavridou, and J.~Schumann, ``Generation of {{Formal Requirements}} from {{Structured Natural Language}},'' in \emph{Requirements {{Engineering}}: {{Foundation}} for {{Software Quality}}}, N.~Madhavji, L.~Pasquale, A.~Ferrari, and S.~Gnesi, Eds.\hskip 1em plus 0.5em minus 0.4em\relax Springer International Publishing, 2020, pp. 19--35.

\bibitem{berryAmbiguityRequirementsSpecification2004}
D.~M. Berry and E.~Kamsties, ``Ambiguity in {{Requirements Specification}},'' in \emph{Perspectives on {{Software Requirements}}}.\hskip 1em plus 0.5em minus 0.4em\relax Springer US, 2004, pp. 7--44.

\bibitem{de2008z3}
L.~De~Moura and N.~Bj{\o}rner, ``{Z3}: An efficient {SMT} solver,'' in \emph{TACAS 2008}.\hskip 1em plus 0.5em minus 0.4em\relax Springer, 2008, pp. 337--340.

\bibitem{chenDecisionProceduresPath2019}
T.~Chen, M.~Hague, A.~W. Lin, P.~R{\"u}mmer, and Z.~Wu, ``Decision procedures for path feasibility of string-manipulating programs with complex operations,'' \emph{OSTRICH String Constraint Solver and Results}, vol.~3, no. POPL, pp. 49:1--49:30, Jan. 2019.

\bibitem{huang2023towards}
J.~Huang and K.~C.-C. Chang, ``Towards reasoning in large language models: A survey,'' in \emph{ACL 2023}.\hskip 1em plus 0.5em minus 0.4em\relax ACL, 2023, pp. 1049--1065.

\bibitem{babikian2025exploring}
\BIBentryALTinterwordspacing
A.~A. Babikian, B.~Chen, and G.~Mussbacher, ``Exploring large language models for requirements on string values,'' in \emph{Proceedings of the 2nd IEEE/ACM Workshop on Multi-disciplinary, Open, and RElevant Requirements Engineering}, 2025. [Online]. Available: \url{https://mo2re.github.io/assets/preprints/BCM-LLMsForStringRequirments-MO2RE25.pdf}
\BIBentrySTDinterwordspacing

\bibitem{li2024guiding}
Y.~Li, J.~Parsert, and E.~Polgreen, ``Guiding enumerative program synthesis with large language models,'' in \emph{International Conference on Computer Aided Verification}.\hskip 1em plus 0.5em minus 0.4em\relax Springer, 2024, pp. 280--301.

\bibitem{first2023baldur}
E.~First, M.~N. Rabe, T.~Ringer, and Y.~Brun, ``Baldur: Whole-proof generation and repair with large language models,'' in \emph{Proceedings of the 31st ACM Joint European Software Engineering Conference and Symposium on the Foundations of Software Engineering}, 2023, pp. 1229--1241.

\bibitem{kirchner2024prover}
J.~H. Kirchner, Y.~Chen, H.~Edwards, J.~Leike, N.~McAleese, and Y.~Burda, ``Prover-verifier games improve legibility of llm outputs,'' \emph{arXiv preprint arXiv:2407.13692}, 2024.

\bibitem{hurst2024gpt}
A.~Hurst, A.~Lerer, A.~P. Goucher, A.~Perelman, A.~Ramesh, A.~Clark, A.~Ostrow, A.~Welihinda, A.~Hayes, A.~Radford \emph{et~al.}, ``Gpt-4o system card,'' \emph{arXiv preprint arXiv:2410.21276}, 2024.

\bibitem{grattafiori2024llama}
A.~Grattafiori, A.~Dubey, A.~Jauhri, A.~Pandey, A.~Kadian, A.~Al-Dahle, A.~Letman, A.~Mathur, A.~Schelten, A.~Vaughan \emph{et~al.}, ``The llama 3 herd of models,'' \emph{arXiv preprint arXiv:2407.21783}, 2024.

\bibitem{liu2024deepseek}
A.~Liu, B.~Feng, B.~Xue, B.~Wang, B.~Wu, C.~Lu, C.~Zhao, C.~Deng, C.~Zhang, C.~Ruan \emph{et~al.}, ``Deepseek-v3 technical report,'' \emph{arXiv preprint arXiv:2412.19437}, 2024.

\bibitem{chenSolvingStringConstraints2023}
Y.-F. Chen, D.~Chocholat\'{y}, V.~Havlena, L.~Hol\'{\i}k, O.~Leng\'{a}l, and J.~S\'{\i}\v{c}, ``Solving string constraints with lengths by stabilization,'' \emph{Proceedings of ACM Programming Languages}, vol.~7, no. OOPSLA2, Oct. 2023.

\bibitem{lotzSolvingStringConstraints2023}
K.~Lotz, A.~Goel, B.~Dutertre, B.~{Kiesl-Reiter}, S.~Kong, R.~Majumdar, and D.~Nowotka, ``Solving {{String Constraints Using SAT}},'' in \emph{CAV 2023}.\hskip 1em plus 0.5em minus 0.4em\relax Springer, 2023, pp. 187--208.

\bibitem{wuDecisionProcedureString2024}
H.~Wu, Y.-F. Chen, Z.~Wu, B.~Xia, and N.~Zhan, ``A decision procedure for string constraints with string/integer conversion and flat regular constraints,'' \emph{Acta Informatica}, vol.~61, no.~1, pp. 23--52, Mar. 2024.

\bibitem{kringsConstraintLogicProgramming2020}
S.~Krings, J.~Schmidt, P.~Skowronek, J.~Dunkelau, and D.~Ehmke, ``Towards {{Constraint Logic Programming}} over {{Strings}} for {{Test Data Generation}},'' in \emph{Declarative {{Programming}} and {{Knowledge Management}}}.\hskip 1em plus 0.5em minus 0.4em\relax Springer, 2020, pp. 139--159.

\bibitem{SMTLIBSatisfiabilityModulo}
``{{SMT-LIB The Satisfiability Modulo Theories Library}},'' https://smt-lib.org/benchmarks.shtml.

\bibitem{SMTCOMP2024}
``{{SMT-COMP}} 2024,'' https://smt-comp.github.io/2024/.

\bibitem{barrettSMTLIBStandardVersion2017}
C.~Barrett, P.~Fontaine, and C.~Tinelli, ``The {{SMT-LIB}} standard: {{Version}} 2.6,'' Department of Computer Science, The University of Iowa, Tech. Rep., 2017.

\bibitem{vaswani2017attention}
A.~Vaswani, N.~M. Shazeer, N.~Parmar, J.~Uszkoreit, L.~Jones, A.~N. Gomez, L.~Kaiser, and I.~Polosukhin, ``Attention is all you need,'' in \emph{NIPS}, 2017.

\bibitem{sahoo2024systematic}
P.~Sahoo, A.~K. Singh, S.~Saha, V.~Jain, S.~Mondal, and A.~Chadha, ``A systematic survey of prompt engineering in large language models: Techniques and applications,'' \emph{arXiv preprint arXiv:2402.07927}, 2024.

\bibitem{li2023large}
C.~Li, J.~Wang, Y.~Zhang, K.~Zhu, W.~Hou, J.~Lian, F.~Luo, Q.~Yang, and X.~Xie, ``Large language models understand and can be enhanced by emotional stimuli,'' \emph{arXiv preprint arXiv:2307.11760}, 2023.

\bibitem{kojima2022large}
T.~Kojima, S.~S. Gu, M.~Reid, Y.~Matsuo, and Y.~Iwasawa, ``Large language models are zero-shot reasoners,'' \emph{NeurIPS 2022}, pp. 22\,199--22\,213, 2022.

\bibitem{openai2025introducing}
\BIBentryALTinterwordspacing
OpenAI, accessed: 2025-03-10. [Online]. Available: \url{https://openai.com/index/introducing-structured-outputs-in-the-api}
\BIBentrySTDinterwordspacing

\bibitem{harrison2022langchain}
\BIBentryALTinterwordspacing
H.~Chase and contributors, ``{LangChain},'' 2022. [Online]. Available: \url{https://github.com/langchain-ai/langchain}
\BIBentrySTDinterwordspacing

\bibitem{dinges2024solving}
P.~Dinges and G.~A. Agha, ``Solving complex path conditions through heuristic search on induced polytopes,'' in \emph{Proceedings of the 22nd {ACM} {SIGSOFT} International Symposium on Foundations of Software Engineering, (FSE-22), Hong Kong, China, November 16 - 22, 2014}.\hskip 1em plus 0.5em minus 0.4em\relax {ACM}, 2014, pp. 425--436.

\bibitem{kandel2020batchsize}
\BIBentryALTinterwordspacing
I.~Kandel and M.~Castelli, ``The effect of batch size on the generalizability of the convolutional neural networks on a histopathology dataset,'' \emph{ICT Express}, vol.~6, no.~4, pp. 312--315, 2020. [Online]. Available: \url{https://www.sciencedirect.com/science/article/pii/S2405959519303455}
\BIBentrySTDinterwordspacing

\bibitem{chen2021evaluating}
M.~Chen, J.~Tworek, H.~Jun, Q.~Yuan \emph{et~al.}, ``Evaluating large language models trained on code,'' \emph{arXiv preprint arXiv:2107.03374}, 2021.

\bibitem{wei2022chain}
J.~Wei, X.~Wang, D.~Schuurmans, M.~Bosma, F.~Xia, E.~Chi, Q.~V. Le, D.~Zhou \emph{et~al.}, ``Chain-of-thought prompting elicits reasoning in large language models,'' \emph{Advances in neural information processing systems}, vol.~35, pp. 24\,824--24\,837, 2022.

\bibitem{2025copilot}
``Github copilot,'' \url{https://github.com/features/copilot}, accessed: 2025-03-10.

\bibitem{Cvc5SrcTheory}
``Cvc5/src/theory/incomplete\_id.h at main {$\cdot$} cvc5/cvc5,'' \url{https://github.com/cvc5/cvc5/blob/main/src/theory/incomplete\_id.h}.

\bibitem{burnstein2006practical}
I.~Burnstein, \emph{Practical software testing: a process-oriented approach}.\hskip 1em plus 0.5em minus 0.4em\relax Springer Science \& Business Media, 2006.

\bibitem{artifact_re2025}
\BIBentryALTinterwordspacing
B.~Chen, A.~A. Babikian, S.~Feng, D.~Varr\'o, and G.~Mussbacher, ``Constrainsolver artifacts,'' Jun. 2025. [Online]. Available: \url{https://doi.org/10.5281/zenodo.15679384}
\BIBentrySTDinterwordspacing

\bibitem{chen2024survey}
L.~Chen, Q.~Guo, H.~Jia, Z.~Zeng, X.~Wang, Y.~Xu, J.~Wu, Y.~Wang, Q.~Gao, J.~Wang \emph{et~al.}, ``A survey on evaluating large language models in code generation tasks,'' \emph{arXiv preprint arXiv:2408.16498}, 2024.

\bibitem{arora2024advancing}
C.~Arora, J.~Grundy, and M.~Abdelrazek, ``Advancing requirements engineering through generative {AI}: Assessing the role of {LLMs},'' in \emph{Generative AI for Effective Software Development}.\hskip 1em plus 0.5em minus 0.4em\relax Springer, 2024, pp. 129--148.

\bibitem{ronanki2023investigating}
K.~Ronanki, C.~Berger, and J.~Horkoff, ``Investigating {ChatGPT}’s potential to assist in requirements elicitation processes,'' in \emph{2023 49th Euromicro Conference on Software Engineering and Advanced Applications (SEAA)}.\hskip 1em plus 0.5em minus 0.4em\relax IEEE, 2023, pp. 354--361.

\bibitem{gorer2023generating}
B.~G{\"o}rer and F.~B. Aydemir, ``Generating requirements elicitation interview scripts with large language models,'' in \emph{RE 2023 Workshops (REW)}.\hskip 1em plus 0.5em minus 0.4em\relax IEEE, 2023, pp. 44--51.

\bibitem{chen2023automated}
K.~Chen, Y.~Yang, B.~Chen, J.~A.~H. L{\'o}pez, G.~Mussbacher, and D.~Varr{\'o}, ``Automated domain modeling with large language models: A comparative study,'' in \emph{2023 ACM/IEEE 26th International Conference on Model Driven Engineering Languages and Systems (MODELS)}.\hskip 1em plus 0.5em minus 0.4em\relax IEEE, 2023, pp. 162--172.

\bibitem{camara2023assessment}
J.~C{\'a}mara, J.~Troya, L.~Burgue{\~n}o, and A.~Vallecillo, ``On the assessment of generative ai in modeling tasks: an experience report with chatgpt and uml,'' \emph{Software and Systems Modeling}, vol.~22, no.~3, pp. 781--793, 2023.

\bibitem{chen2023use}
B.~Chen, K.~Chen, S.~Hassani, Y.~Yang, D.~Amyot, L.~Lessard, G.~Mussbacher, M.~Sabetzadeh, and D.~Varr{\'o}, ``On the use of {GPT-4} for creating goal models: An exploratory study,'' in \emph{RE 2023 Workshops}.\hskip 1em plus 0.5em minus 0.4em\relax IEEE, 2023, pp. 262--271.

\bibitem{de2024towards}
S.~de~Kinderen and K.~Winter, ``Towards taming large language models with prompt templates for legal {GRL} modeling,'' in \emph{BPMDS 2024}.\hskip 1em plus 0.5em minus 0.4em\relax Springer, 2024, pp. 213--228.

\bibitem{liu2023fill}
Z.~Liu, C.~Chen, J.~Wang, X.~Che, Y.~Huang, J.~Hu, and Q.~Wang, ``Fill in the blank: Context-aware automated text input generation for mobile {GUI} testing,'' in \emph{ICSE 2023}.\hskip 1em plus 0.5em minus 0.4em\relax IEEE, 2023, pp. 1355--1367.

\bibitem{xia2024fuzz4all}
C.~S. Xia, M.~Paltenghi, J.~Le~Tian, M.~Pradel, and L.~Zhang, ``Fuzz4all: Universal fuzzing with large language models,'' in \emph{ICSE 2024}, 2024, pp. 1--13.

\bibitem{sun2023smt}
M.~Sun, Y.~Yang, Y.~Wang, M.~Wen, H.~Jia, and Y.~Zhou, ``{SMT} solver validation empowered by large pre-trained language models,'' in \emph{ASE 2023}.\hskip 1em plus 0.5em minus 0.4em\relax IEEE, 2023, pp. 1288--1300.

\bibitem{liu2024testing}
Z.~Liu, C.~Chen, J.~Wang, M.~Chen, B.~Wu, Z.~Tian, Y.~Huang, J.~Hu, and Q.~Wang, ``Testing the limits: Unusual text inputs generation for mobile app crash detection with large language model,'' in \emph{ICSE 2024}, 2024, pp. 1--12.

\bibitem{yao2023tree}
S.~Yao, D.~Yu, J.~Zhao, I.~Shafran, T.~Griffiths, Y.~Cao, and K.~Narasimhan, ``Tree of thoughts: Deliberate problem solving with large language models,'' \emph{Advances in neural information processing systems}, vol.~36, pp. 11\,809--11\,822, 2023.

\bibitem{long2023large}
J.~Long, ``Large language model guided tree-of-thought,'' \emph{arXiv preprint arXiv:2305.08291}, 2023.

\bibitem{besta2024graph}
M.~Besta, N.~Blach, A.~Kubicek, R.~Gerstenberger, M.~Podstawski, L.~Gianinazzi, J.~Gajda, T.~Lehmann, H.~Niewiadomski, P.~Nyczyk \emph{et~al.}, ``Graph of thoughts: Solving elaborate problems with large language models,'' in \emph{Proceedings of the AAAI Conference on Artificial Intelligence}, vol.~38, no.~16, 2024, pp. 17\,682--17\,690.

\bibitem{hong2023metagpt}
S.~Hong, X.~Zheng, J.~Chen, Y.~Cheng, J.~Wang, C.~Zhang, Z.~Wang, S.~K.~S. Yau, Z.~Lin, L.~Zhou \emph{et~al.}, ``Metagpt: Meta programming for multi-agent collaborative framework,'' \emph{arXiv preprint arXiv:2308.00352}, vol.~3, no.~4, p.~6, 2023.

\bibitem{li2024more}
J.~Li, Q.~Zhang, Y.~Yu, Q.~Fu, and D.~Ye, ``More agents is all you need,'' \emph{arXiv preprint arXiv:2402.05120}, 2024.

\bibitem{wu2023autogen}
Q.~Wu, G.~Bansal, J.~Zhang, Y.~Wu, B.~Li, E.~Zhu, L.~Jiang, X.~Zhang, S.~Zhang, J.~Liu \emph{et~al.}, ``Autogen: Enabling next-gen {LLM} applications via multi-agent conversation,'' \emph{arXiv preprint arXiv:2308.08155}, 2023.

\bibitem{gao2024agentscope}
D.~Gao, Z.~Li, X.~Pan, W.~Kuang, Z.~Ma, B.~Qian, F.~Wei, W.~Zhang, Y.~Xie, D.~Chen \emph{et~al.}, ``Agentscope: A flexible yet robust multi-agent platform,'' \emph{arXiv preprint arXiv:2402.14034}, 2024.

\bibitem{he2024llm}
J.~He, C.~Treude, and D.~Lo, ``Llm-based multi-agent systems for software engineering: Literature review, vision and the road ahead,'' \emph{ACM Transactions on Software Engineering and Methodology}, 2024.

\bibitem{alshahwan2024automated}
N.~Alshahwan, J.~Chheda, A.~Finogenova, B.~Gokkaya, M.~Harman, I.~Harper, A.~Marginean, S.~Sengupta, and E.~Wang, ``Automated unit test improvement using large language models at {Meta},'' in \emph{Companion Proceedings of the 32nd ACM International Conference on the Foundations of Software Engineering}, 2024, pp. 185--196.

\bibitem{yang2024multi}
Y.~Yang, B.~Chen, K.~Chen, G.~Mussbacher, and D.~Varro, ``Multi-step iterative automated domain modeling with large language models,'' in \emph{ACM/IEEE 27th International Conference on Model Driven Engineering Languages and Systems: Companion Proceedings (MODELS)}, 2024, pp. 587--595.

\bibitem{masoudifard2024leveraging}
A.~Masoudifard, M.~M. Sorond, M.~Madadi, M.~Sabokrou, and E.~Habibi, ``Leveraging graph-rag and prompt engineering to enhance llm-based automated requirement traceability and compliance checks,'' \emph{arXiv preprint arXiv:2412.08593}, 2024.

\bibitem{babikianConcretizationAbstractTraffic2024}
A.~A. Babikian, O.~Semeráth, and D.~Varró, ``Concretization of {{Abstract Traffic Scene Specifications Using Metaheuristic Search}},'' \emph{IEEE Transactions on Software Engineering}, vol.~50, no.~1, pp. 48--68, 2024.

\end{thebibliography}

\end{document}